\documentclass{emulateapj_accepted}

\usepackage{rotating,natbib}
\bibstyle{apj}

\def \hii{\rm H{ \scriptsize II}}

\def \water{\mbox{H$_2$O}}

\def \methanol{\mbox{CH$_3$OH}}
\def \methylcyanide{\mbox{CH$_3$CN}}
\def \twelveco{\mbox{$^{12}$CO}}
\def \twelvecotto{\mbox{$^{12}$CO\,(2$\rightarrow$1)}}
\def \thirteenco{\mbox{$^{13}$CO}}
\def \thirteencotto{\mbox{$^{13}$CO\,(2$\rightarrow$1)}}

\def \nhthree{\mbox{NH$_3$}}
\def \nhone{\mbox{NH$_3$(1,1)}}

\def \nhfour{\mbox{NH$_3$(4,4)}}

\def \kms{\mbox{kms$^{-1}$}}
\def \vlsr{\mbox{V$_{\rm LSR}$}}
\def \arcsec{\mbox{$^{\prime\prime}$}}
\def \twotoone{\mbox{(2$\rightarrow$1)}}

\def \cm2{\mbox{cm$^{-2}$}}
\def \cm3{\mbox{cm$^{-3}$}}
\def\lsim{\mathrel{\rlap{\lower4pt\hbox{\hskip1pt$\sim$}}\raise1pt\hbox{$<$}}} 
\def\gsim{\mathrel{\rlap{\lower4pt\hbox{\hskip1pt$\sim$}}\raise1pt\hbox{$>$}}}  

\shorttitle{Fragmentation \& protostellar heating in a massive protocluster}
\shortauthors{S. N. Longmore et al.}

\begin{document}

\title{Is protostellar heating sufficient to halt fragmentation? A
  case study of the massive protocluster G8.68$-$0.37}

\author{S. N. Longmore\altaffilmark{1}, T. Pillai,
E. Keto, Q. Zhang \& K. Qiu} 

\affil{Harvard-Smithsonian Center for Astrophysics, 60 Garden Street,
Cambridge, MA 02138}

\altaffiltext{1}{Contact email: slongmore@cfa.harvard.edu}

\begin{abstract}

If star formation proceeds by thermal fragmentation and the subsequent
gravitational collapse of the individual fragments, how is it possible
to form fragments massive enough for O and B stars in a typical
star-forming molecular cloud where the Jeans mass is about
1\,M$_\odot$ at the typical densities (10$^4$\,$\cm3$) and
temperatures (10\,K)?  We test the hypothesis that a first generation
of low-mass stars may heat the gas enough that subsequent thermal
fragmentation results in fragments $\geq$10M$_\odot$, sufficient to
form B stars. We combine ATCA and SMA observations of the massive
star-forming region G8.68$-$0.37 with radiative transfer modeling to
derive the present-day conditions in the region and use this to infer
the conditions in the past, at the time of core formation. Assuming
the current mass/separation of the observed cores equals the
fragmentation Jeans mass/length and the region's average density has
not changed, requires the gas temperature to have been 100\,K at the
time of fragmentation. The postulated first-generation of low-mass
stars would still be around today, but the number required to heat the
cloud exceeds the limits imposed by the observations. Several lines of
evidence suggest the observed cores in the region should eventually
form O stars yet none have sufficient raw material. Even if feedback
may have suppressed fragmentation, it was not sufficient to halt it to
this extent. To develop into O stars, the cores must obtain additional
mass from outside their observationally defined boundaries. The
observations suggest they are currently fed via infall from the very
massive reservoir ($\sim$1500\,M$_\odot$) of gas in the larger pc
scale cloud around the star-forming cores.  This suggests that massive
stars do not form in the collapse of individual massive fragments, but
rather in smaller fragments that themselves continue to gain mass by
accretion from larger scales.

\end{abstract}


\section{Introduction}
\label{sec:intro}

Massive stars play an influential role in shaping the
Universe. Observations show that the substantial majority form in
clusters \citep{ladalada2003} yet the physical processes governing the
fragmentation and collapse of their natal molecular cloud, a crucial
step in determining important parameters such as the number of massive
stars and their final stellar mass, remains an unsolved
problem. Infrared dark clouds, thought to be examples of the birth
sites of massive clusters, are observed to have temperatures of
$10-20$\,K and contain many hundreds to thousands of solar masses of
gas \citep{pillai2006, ragan2006, rathborne2006, swift2009, zhang2009,
  rathborne2010}. It remains an open question whether these initial
conditions can produce fragments large enough to form high-mass stars
through direct collapse, without the cores themselves sub-fragmenting
first. Radiative feedback (heating) from embedded, low mass
protocluster members has been proposed as a mechanism to delay
fragmentation by changing the effective equation of state
\citep{krumholz2007}. In this paper we consider a more direct
mechanism to suppress the sub-fragmentation. Through observations of a
very young high-mass protocluster, we investigate whether heating from
a first generation of low mass stars can raise the cloud temperature
enough that subsequent thermal fragmentation can produce fragments
with sufficient mass to form high-mass stars.

\subsection{The IRAS 18032$-$2137 complex}
\label{sub:iras18032}
\begin{figure*}
\begin{center}
 \includegraphics[width=14cm, angle=0, trim=0 0 -5 0]{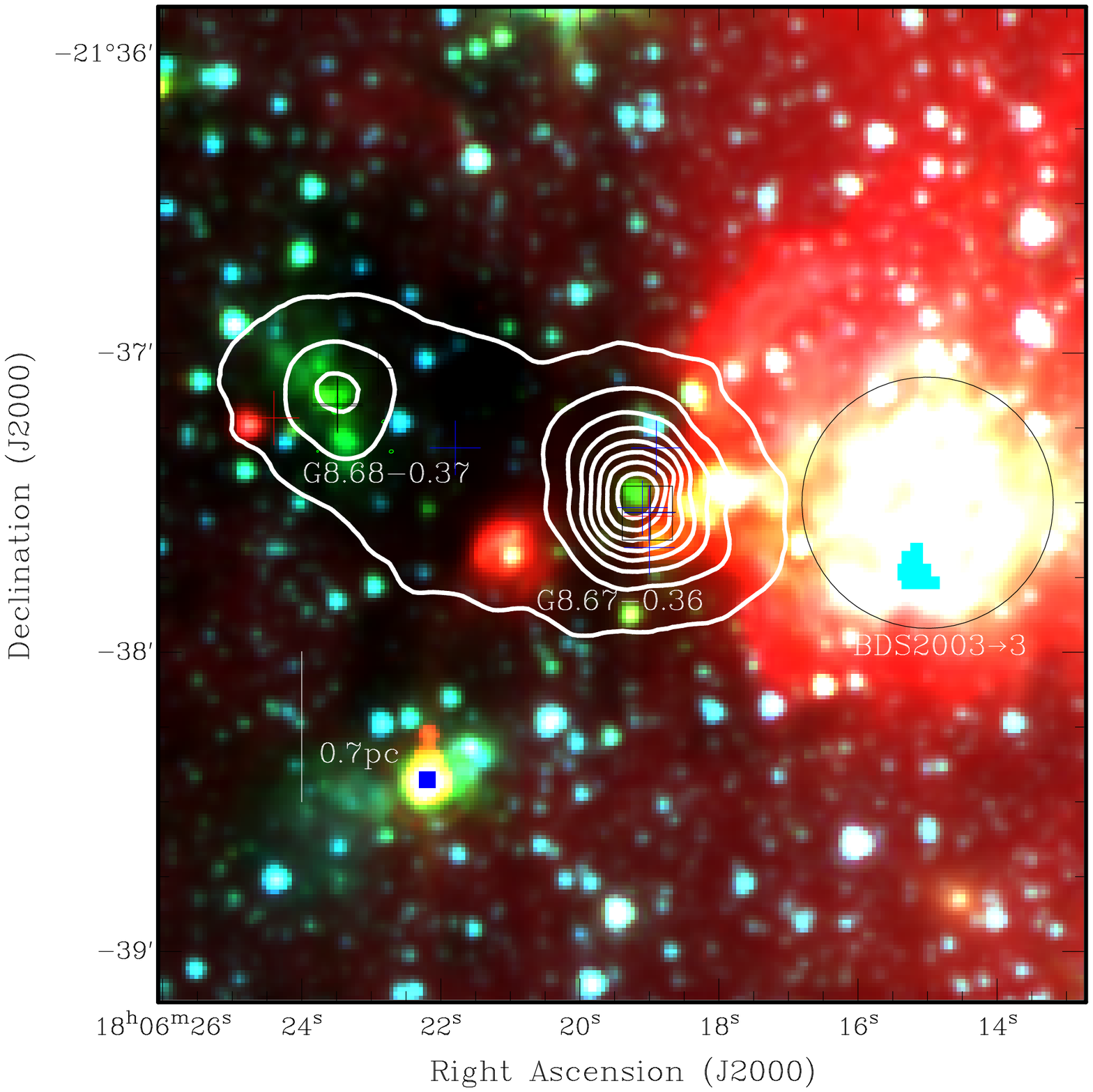}\\
\end{center}
\caption{The IRAS 18032$-$2137 star forming complex. A GLIMPSE
  3-colour image at 3.6$\mu$m, 4.5$\mu$m and 8.0$\mu$m is overlayed
  with SCUBA 850\,$\mu$m continuum emission contours
  \citep{hill2006}. The complex comprises 3 distinct regions: i) the
  stellar cluster, ${\rm BDS2003-3}$ \citep{bica2003}, with the
  approximate extent reported by \citet{bica2003} indicated by the
  circle; ii) the ultra-compact $\hii$ region G$8.67-0.36$
  \citep[][cm-continuum position illustrated by box ]{WC1989}; iii) a
  massive star forming core prior to formation of an HII region,
  G$8.68-0.37$. $\water$, $\methanol$ and OH maser emission
  \citep{hofner_churchwell1996, walsh1998, forster_caswell1989,
    caswell1998, valtts2000} are shown as blue, black and red crosses,
  respectively. The line towards the lower left gives the physical
  scale for the source distance of 4.8\,kpc \citep{purcell2006}.}
\label{fig:g8.68_large_scale_3col_glimpse}
\end{figure*}

Figure~\ref{fig:g8.68_large_scale_3col_glimpse} shows the IRAS
18032$-$2137 star forming complex, comprised of 3 distinct regions
separated by a few arcminutes on the sky. The most evolved of these is
the stellar cluster, BDS2003-3 \citep[$\alpha_{\rm J2000}$=18:06:15,
  $\delta_{\rm J2000} =$ -21:37:30,][]{bica2003}. The high visual
extinction, association with both near-IR nebulosity \citep{L09_I2}
and radio continuum emission \citep[G8.662-0.342,][]{becker1994} make
this most likely a massive, heavily embedded cluster. One arcminute
east of the cluster lies the ultra-compact $\hii$ region G$8.67-0.36$
\citep{WC1989}, coincident with a compact sub-mm continuum peak
\citep{hill2005,hill2006,thompson2006}. One arcminute (projected
separation of 1.4\,pc) north-east along the same sub-mm continuum
filament lies a second sub-mm continuum peak, G$8.68-0.37$. Both
G$8.67-0.36$ and G$8.68-0.37$ are associated with $\water$, class II
$\methanol$ and OH maser emission \citep{hofner_churchwell1996,
  walsh1998, forster_caswell1989, caswell1998,
  valtts2000}. \citet{hill2010} show G$8.68-0.37$ is massive
($\sim$1.5$\times$10$^3$\,M$_\odot$) and has a high luminosity
($\sim$10$^4$L$_\odot$) yet deep cm-continuum observations reveal no
compact (Michele Pestallozi priv. comm.) or extended
\citep{longmore2009} free-free emission. At a distance of 4.8\,kpc
\citep{purcell2006}, the 3$\sigma$ upper limit of 1.7mJy at 3.6\,cm
(Michele Pestallozi priv. comm.) corresponds to a Lyman-continuum
photon rate of 1.65$\times$10$^{46}$\,s$^{-1}$, implying no stars
earlier than B0-B1 are present in the cluster. All of this points to
the nature of G$8.68-0.37$ as a very young massive star forming
region, prior to the formation of ultra-compact or hyper-compact HII
regions.

$\nhthree$ observations of G$8.68-0.37$ show: (i) an extended cold
component seen in the $\nhone$ \& (2,2) transitions with a morphology
similar to the sub-mm continuum emission; (ii) a warmer component in
$\nhfour$ \& (5,5), unresolved at 8$\arcsec$ at the peak of the sub-mm
continuum and methanol maser emission \citep{pillai2007, L07A}. This
suggests G$8.68-0.37$ is internally heated by young (proto)stars of at
least several M$_\odot$. Strong infall profiles are seen in 3mm Mopra
spectra of HCO$^+$, HNC and even $^{13}$CO \citep{purcell2006,
  purcell2009}, while \citet{harju1998} report strong SiO emission
indicative of shocks caused by outflows -- further evidence that star
formation is already underway in the protocluster.

In summary, G$8.68-0.37$ is a relatively isolated massive star forming
core with a symmetric and centrally peaked dust profile. It is still
in the earliest stages of forming a massive protocluster and has
already begun to significantly heat the gas in the proto-cluster
center. As such it appears a good candidate region for testing whether
energetic feedback can act sufficiently quickly to suppress thermal
fragmentation.

\section{Observations and data reduction}
\label{sec:obs_dr}

The data were taken with the Submillimeter Array\footnote{The
  Submillimeter Array is a joint project between the Smithsonian
  Astrophysical Observatory and the Academia Sinica Institute of
  Astronomy and Astrophysics and is funded by the Smithsonian
  Institution and the Academia Sinica \citep{ho2004}.}  (SMA) between
2007 September 2$^{nd}$ and 2008 September 17$^{th}$ in three
individual tracks between 217 and 279\,GHz in the sub-compact, compact
and extended array configurations. At these frequencies, the SMA
primary beam (field-of-view) is $\sim 45 \arcsec - 58 \arcsec$,
sufficient to cover the extent of the single-dish continuum emission
(see Figure~\ref{fig:g8.68_large_scale_3col_glimpse}) in a single
pointing centered at $\alpha_{J2000}=$18:06:23.47,
$\delta_{J2000}=-$21:37:7.6.  For each track, this sky position was
observed for periods of 10-15 minutes on-source, interspersed with
observations of two bright, nearby calibrators (1733$-$130 \&
1911$-$201). Two bandpass calibrators, 3c279 and 3c454.3, were
observed for approximately an hour at the start and end of each
observation. At least one primary flux calibrator (Uranus, Neptune,
Titan) was observed for each track. The absolute flux scale is
estimated to be accurate to $\sim$15\%. In all cases the weather was
very stable and the resulting amplitude/phase stability was good. The
data were calibrated using the MIR IDL
package\footnote{http://www.cfa.harvard.edu/$\sim$cqi/mircook.html}
and exported to Miriad to be imaged and cleaned. A zeroeth order
polynomial was fitted to the line-free channels and subtracted from
the visibilities to separate the line and continuum emission, which
were imaged separately. Table~\ref{tab:obs_setup} lists the observing
setup, continuum sensitivity and resolution for each of the
tracks. The results are shown in $\S$~\ref{sub:sma_data}.

\section{Results}
\label{sec:results}

\subsection{SMA Data}
\label{sub:sma_data}

In the sub-compact and compact configuration data, the continuum
morphology is similar to that seen in the single-dish observations --
a single component, centrally peaked towards the methanol maser
position. However, as shown in Figure~\ref{fig:g8.68_rgb}, the SMA
extended configuration observations resolve the emission into 3
components, named MM1 to MM3 in order of peak intensity. These are
separated by between 1.4\arcsec and 2.2\arcsec, corresponding to
projected distances of $\sim$6200\,AU and $\sim$9700\,AU,
respectively. Table~\ref{tab:mm_src_props} gives the properties of the
mm continuum detections.

\begin{figure*}
\begin{center}
 \includegraphics[width=12cm, angle=-90, trim=0 0 -5 0]{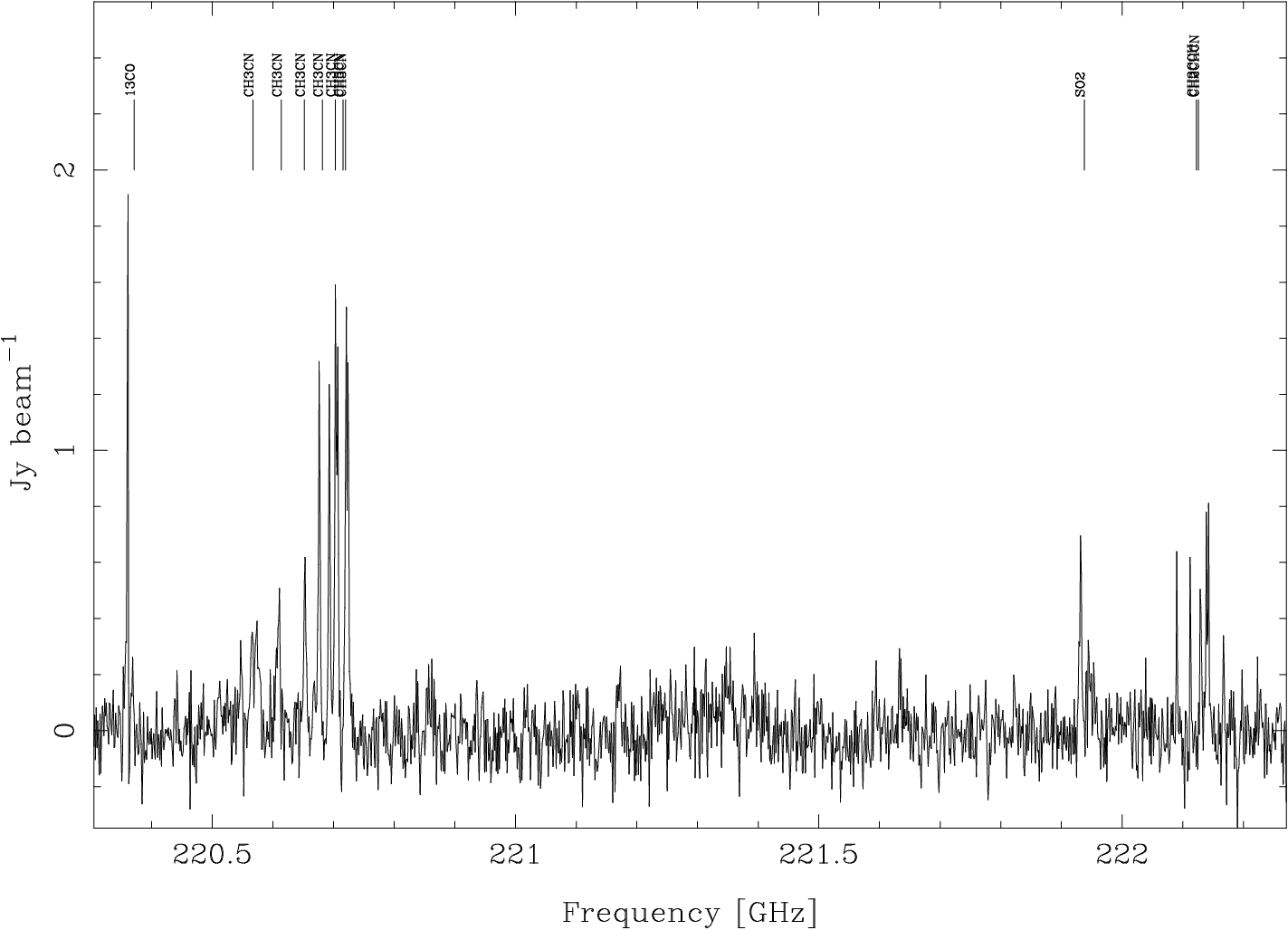}\\
\end{center}
\caption{Lower sideband spectrum of the 230\,GHz,
  compact-configuration SMA data taken at the position of peak
  continuum emission and smoothed to a resolution of
  1.75\,km/s. Transitions with 2 consecutive channels $>$3$\sigma$
  ($\sigma = 92$\,mJy/channel) are labelled above the spectrum at
  frequencies (corrected for the source V$_{\rm LSR}$ of 37.2\,$\kms$)
  taken from the NIST catalog \citep{lovas2004}. In addition to the
  $\thirteencotto$ and $\twelvecotto$ emission, the spectra show
  several more complex molecules (e.g. $\methanol$ \&
  $\methylcyanide$) confirming the source is warm enough for these
  molecules to have evaporated off the dust grains and into the gas
  phase. However, the spectra do not display the rich inventory seen
  towards hot molecular cores suggesting this region is an
  intermediate evolutionary stage between the cold and hot core
  stages.}
\label{fig:g8.68_full_spectra_lsb}
\end{figure*}

\begin{figure*}
\begin{center}
 \includegraphics[width=12cm, angle=-90, trim=0 0 -5 0]{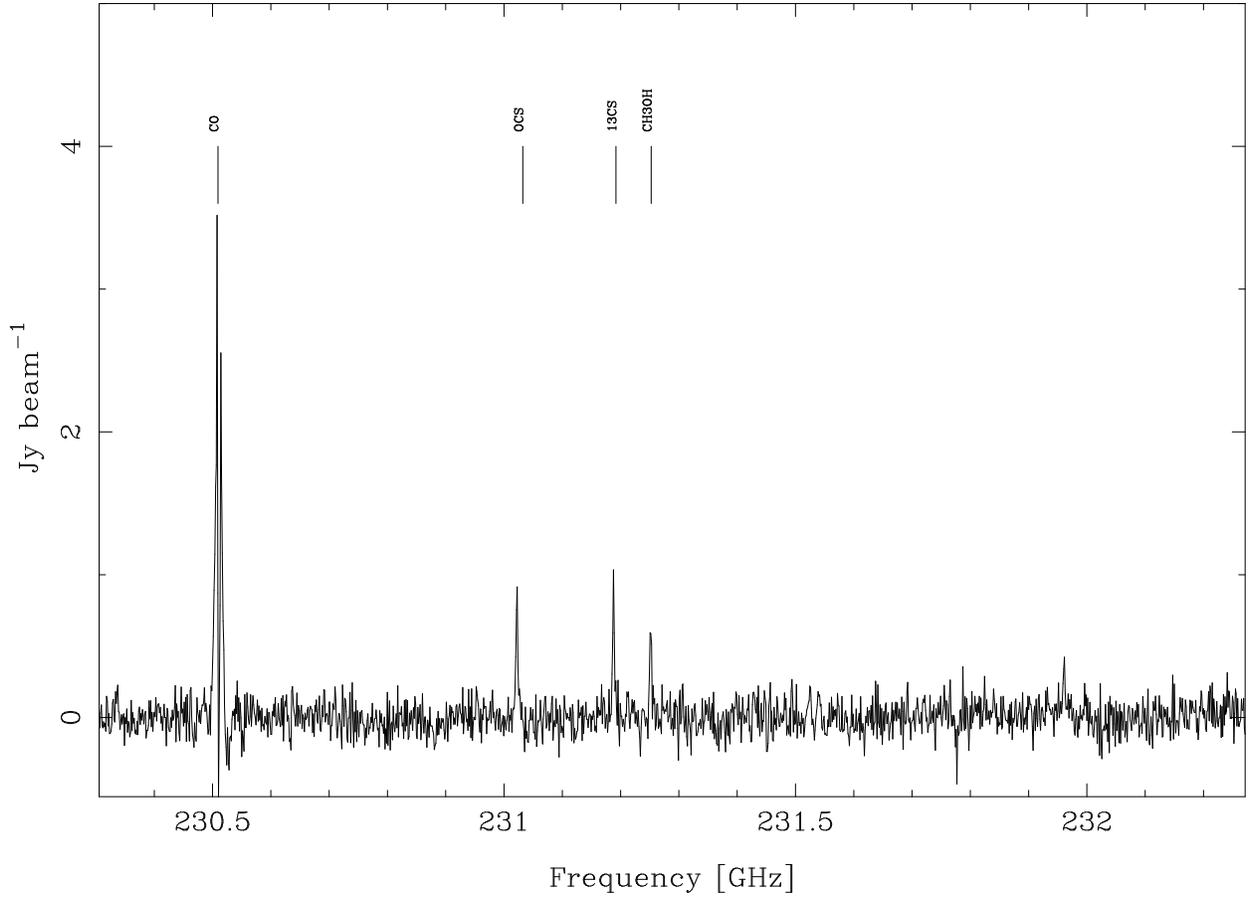}\\
\end{center}
\caption{Upper sideband spectrum of the 230\,GHz,
  compact-configuration SMA data taken at the position of peak
  continuum emission and smoothed to a resolution of
  1.75\,km/s. Transitions with 2 consecutive channels $>$3$\sigma$
  ($\sigma = 94$\,mJy/channel) are labelled above the spectrum at
  frequencies (corrected for the source V$_{\rm LSR}$ of 37.2\,$\kms$)
  taken from the NIST catalog \citep{lovas2004}. In addition to the
  $\thirteencotto$ and $\twelvecotto$ emission, the spectra show
  several more complex molecules (e.g. $\methanol$ \&
  $\methylcyanide$) confirming the source is warm enough for these
  molecules to have evaporated off the dust grains and into the gas
  phase. However, the spectra do not display the rich inventory seen
  towards hot molecular cores suggesting this region is an
  intermediate evolutionary stage between the cold and hot core
  stages.}
\label{fig:g8.68_full_spectra_usb}
\end{figure*}

Figures~\ref{fig:g8.68_full_spectra_lsb}~\&~\ref{fig:g8.68_full_spectra_usb}
show the 230\,GHz lower and upper side-band spectra, respectively,
towards the peak of the continuum emission from the compact
configuration data. In addition to the $\thirteencotto$ and
$\twelvecotto$ emission, the spectra show several more complex
molecules (e.g. $\methanol$ \& $\methylcyanide$) confirming the
temperature is sufficiently high for these molecules to have
evaporated off the dust grains and into the gas phase. However, the
spectra do not display the rich inventory of molecular lines seen
towards hot molecular cores \citep[see][for a review]{cesaroni2005}
suggesting this region is an intermediate evolutionary stage between
the cold and hot core stages. A similar result is found for the
extended configuration observations which were tuned to 217 GHz
(instead of 230 GHz). Figure~\ref{fig:g8.68_12co_chmap} shows the
$\twelvecotto$ channel maps in which a bipolar morphology is clearly
seen at opposite sides of the systemic $\vlsr$ (37.2\,$\kms$),
indicative of a molecular outflow.  The properties of this outflow are
discussed in more detail in $\S$~\ref{sub:outflow}.

The emission from more complex species detected in the 230\,GHz
compact configuration is generally unresolved within the
3\arcsec$\times$2\arcsec\, beam and encompasses MM1 to MM3. This
emission is also coincident with the $\nhfour$ \& (5,5) and methanol
maser emission. The spectral lines are typically well fit as a single
Gaussian at the V$_{\rm LSR}$, with linewidth $\sim$5\kms~and no sign
of any coherent, large-scale velocity structure.

\subsection{Spitzer Data}
\label{sub:spitzer_data}

Figure~\ref{fig:g8.68_rgb} shows a 3-color GLIMPSE image
\citep{benjamin2003} at 3.6, 4.5 and 8.0\,$\mu$m towards the region
taken with IRAC on the Spitzer Space Telescope. There is a clear
bipolar nebulosity in the image, prominent in all four IRAC bands. The
black contours show the positions of the three mm-continuum cores
detected with the SMA relative to the IRAC emission. Of the three mm
cores, the nebulosity is most closely associated with MM1. In the
longer wavelengths (24 \& 70\,$\mu$m) and lower resolution
($7.5-22\arcsec$) MIPSGAL data \citep{carey2009}, only a single,
bright (1 Jy integrated flux at 24\,$\mu$m) emission peak is seen with
MIPS.

\section{Analysis}
\label{sec:analysis}

\subsection{Total Luminosity of G8.68$-$0.37}
While luminosity estimates for this region are available from the
literature, access to recent Spitzer and deep near-IR data not
available to previous authors enables us to provide tighter
constraints. We used the Spitzer 24 \& 70\,$\mu$m, SCUBA 450 \&
850\,$\mu$m and SIMBA 1.3mm (Hill et al. 2005) data to estimate the
total luminosity of the system. Deep, near-IR imaging show this region
is infrared-dark down to 18th magnitude at 2.2$\mu$m \citep{L09_I2},
giving confidence that most of the luminosity is emitted at the longer
wavelengths we used to construct the spectral energy
distribution. Based on the similar size and morphology of the emission
from 24$\mu$m to 1.3mm, we used a circular aperture of radius
15$\arcsec$ to derive the total flux at each wavelength. We fit the
resulting spectral energy distribution as a greybody and, integrating
under the resultant fit, estimate the protocluster luminosity to be
$\sim$1.9$\times$10$^4$\,L$_\odot$. The SMA data were not used in the
luminosity estimate as comparison with the SIMBA data (see
$\S$~\ref{subsub:cont_modelling}) revealed a large fraction of the
total flux has been spatially filtered by the interferometer.

\subsection{Outflow Properties}
\label{sub:outflow}
\begin{figure*}
\begin{center}
 \includegraphics[width=14cm, angle=0, trim=0 0 -5 0]{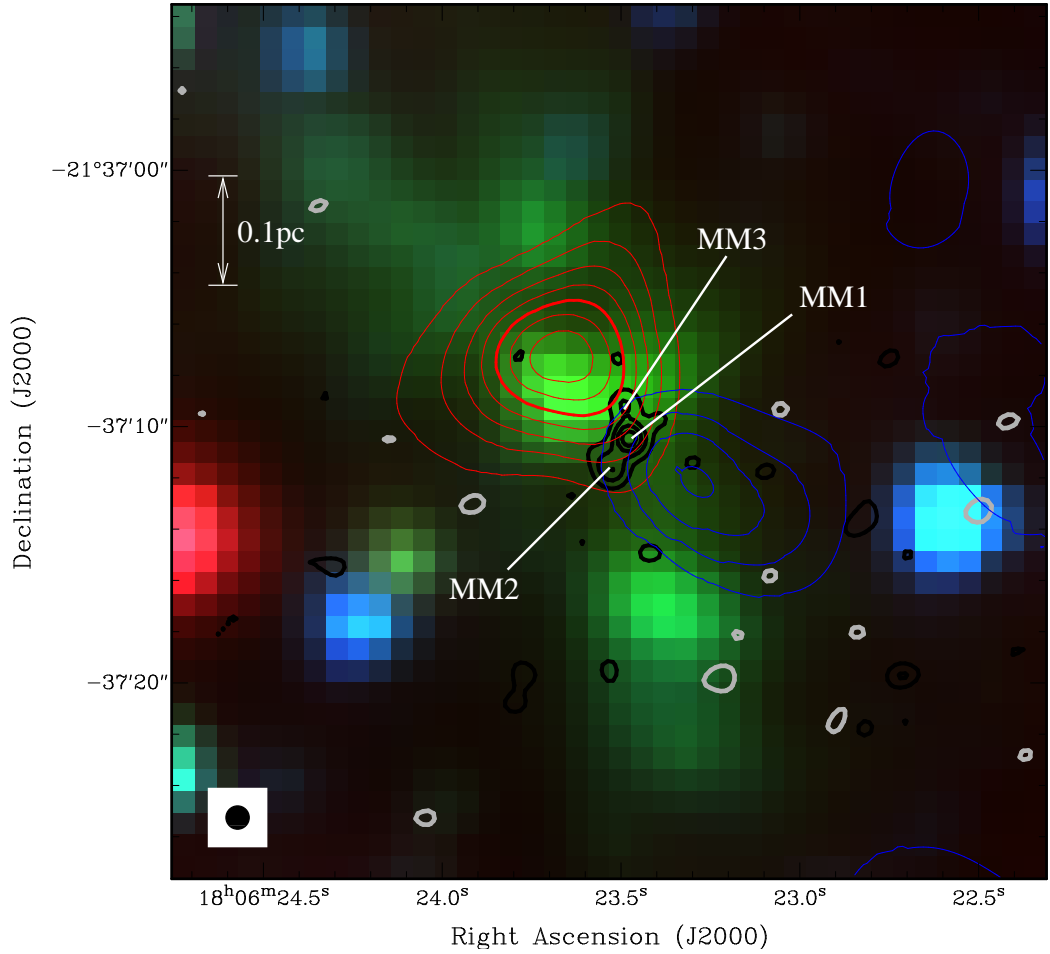}\\
\end{center}
\caption{3-color GLIMPSE image at 3.6, 4.5 and 8.0\,$\mu$m overlayed
  with (i) the 217\,GHz SMA extended-configuration continuum image
  [grey ($-3\sigma$) and black (3, 6, 9$\sigma$, ...) contours:
    $\sigma~=~2.6$\,mJy] and (ii) the $\twelvecotto$ emission
  integrated from 29.3--36.3 km/s [blue contours] and 45--74 km/s [red
    contours]. The blue and red lobes of the molecular outflow lie
  along a similar axis as the extended 4.5\,$\mu$m emission -- a
  well-known tracer of gas shocked in outflows. The 3 SMA 1\,mm
  emission peaks are labelled MM1 to MM3 in order of peak 1\,mm
  continuum emission flux. The brightest component, MM1, is found at
  the center of the outflow lobes and appears the best candidate for
  the outflow driving source. The filled black cirle in the lower-left
  corner gives the synthesized beam of the SMA extended configuration
  observations. The bar in the upper-right corner illustrates the
  physical scale for the source distance of 4.8\,kpc.}
\label{fig:g8.68_rgb}
\end{figure*}

\begin{figure*}
\begin{center}
 \includegraphics[width=14cm, angle=-90, trim=0 0 -5 0]{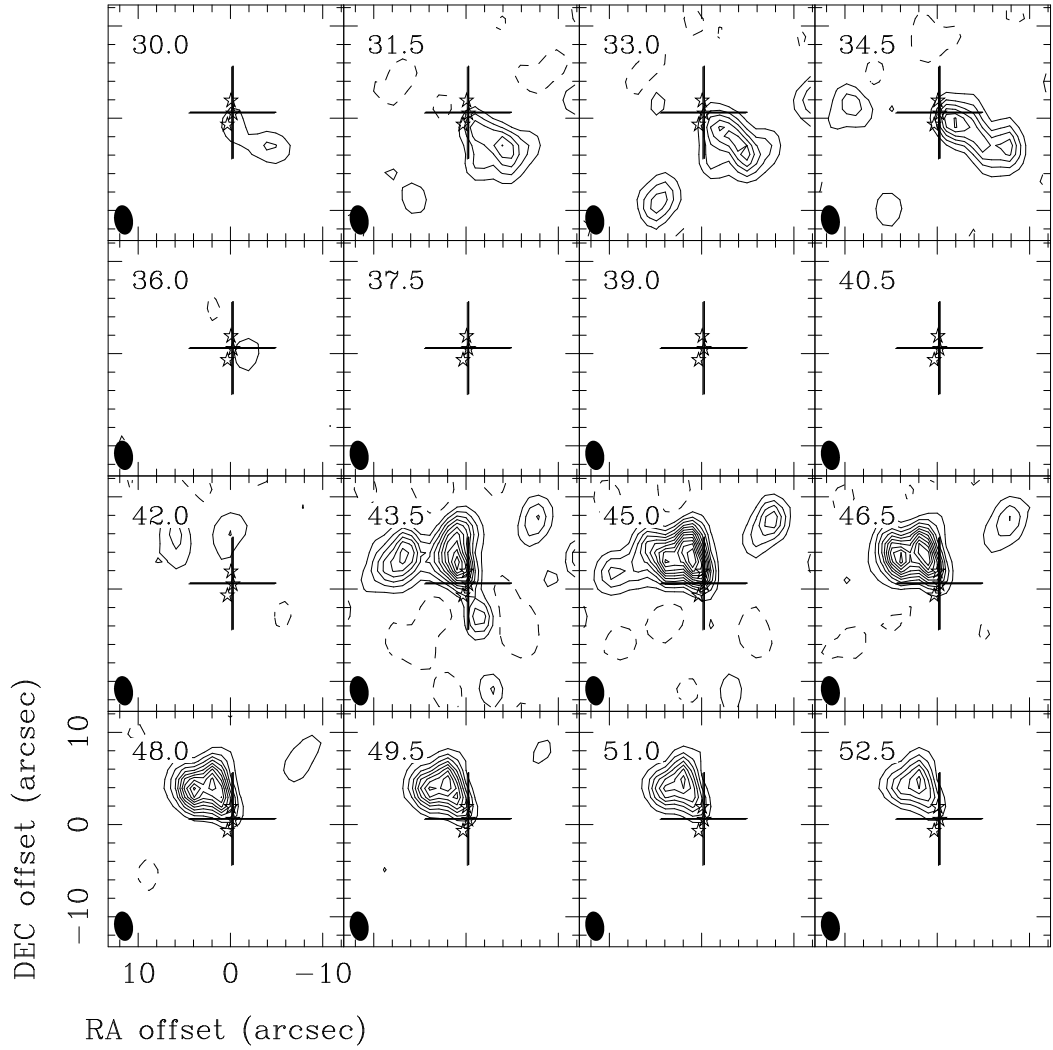}\\
\end{center}
\caption{$\twelveco \twotoone$ channel map towards G$8.68-0.37$. The
  V$_{\rm LSR}$ (in $\kms$) of each frame is shown in the top-left of
  each panel and the synthesised beam size is shown as a filled
  ellipse in the lower left corner of each frame.  The center of the
  coordinate system and location of the cross show the methanol maser
  position at $\alpha_{J2000}~=$~18:06:23.47,
  $\delta_{J2000}~=~-$21:37:10.6. The contours start at $\pm5\sigma$
  and increment in 5$\sigma$ intervals ($\sigma =$
  140\,mJy/beam). For reference, the three stars show the
    location of MM1, MM2 and MM3.}
\label{fig:g8.68_12co_chmap}
\end{figure*}

As shown in Figure~\ref{fig:g8.68_rgb}, both the 230 GHz
compact-configuration, CO SMA data and Spitzer data reveal a bipolar
outflow in an approximately NE-SW orientation. The higher angular
resolution, extended-configuration, 217GHz continuum SMA data shows
core MM1 lies directly along the projected outflow axis of both the CO
and IRAC nebulosity making it most likely to be driving the outflow.
Figure~\ref{fig:g8.68_12co_chmap} shows channel maps of the
$\twelvecotto$ emission.  Extended emission from the ambient cloud is
filtered out by the interferometer (our shortest baseline corresponds
to a spatial filtering of emission more extended than 30$\arcsec$)
close to the LSR velocity of 37.2$\kms$ (Purcell et al.  2006, 2009).
To determine the outflow properties we first separated the outflow
into its two component lobes using a 3 sigma cutoff in the $\twelveco$
data cube at velocities 29.3--36.3 km/s (blue) and 45--74 km/s
(red). Velocities closer to the LSR velocity were avoided to prevent
contamination from the ambient low density gas.

The outflow mass was derived following \citet{scoville1986}, assuming
a CO excitation temperature of 30\,K. We defined a blue and red
polygon covering the spatial extent of the outflow emission shown in
Figure~\ref{fig:g8.68_rgb}, and derived an average spectrum for both
$\twelveco$ \& $\thirteenco$. Where $\thirteenco$ was detected, we
estimated the $\twelveco$ optical depth by assuming the $\thirteenco$
emission to be optically thin with an abundance ratio to $\twelveco$
of 89 \citep[see][]{wilson_matteucci1992}.  The $\twelveco$ optical
depth was then used to compute the CO column density. For velocities
with no $\thirteenco$ emission, optically thin $\twelveco$ emission
was assumed. We thus derive a total outflow mass of
$\sim$6\,M$_\odot$.

Based on the distance between the edge of the lobes through the
projected center, the outflow has a dynamical time of
$\sim$1.3$\times$10$^4$ years. The derived outflow rate is therefore
$\sim$4.4$\times$10$^{-4}$ M$_\odot$ per year. We emphasize that due
to missing flux (see Figure~\ref{fig:g8.68_12co_chmap}) and optical
depth effects, the mass and outflow rate could be
underestimated. Still, for the luminosity of the protocluster, the
outflow properties are in good agreement with those seen towards other
high mass star forming regions \citep[see for
  example,][]{shepherd_churchwell1996, beuther2002_outflows,
  zhang2001, zhang2005, qiu2009}.

Outflow studies in high mass star forming regions are often hampered
by the complexity of the observed emission. In most cases one sees
multiple outflows, many with unknown driving sources, which may be
difficult to disentangle. While there are potentially signs of
additional outflows in Figure~\ref{fig:g8.68_12co_chmap} (e.g. the SE
clump at offset $5\arcsec, - 10\arcsec$ and 31.5 to 34.5\,$\kms$ and
the NW clump at offset $8\arcsec, 8\arcsec$ and 43.5 to 48.0\,$\kms$
may make another outflow), G8.68$-$0.37 is dominated by the simple,
bipolar morphology discussed above which is wide-angled, especially
the red lobe.  As shown in Figure~\ref{fig:g8.68_rgb} the CO outflow
is well aligned with the shocked emission seen in IRAC bands
presumably from the interaction of the outflow with the molecular
envelope.

\citet{cyganowski2009} find methanol masers are detected towards a
large fraction of massive star forming regions associated with
extended 4.5\,$\mu$m emission, so the association of the two in
G8.68$-$0.37 is not surprising. However, the resolution of the SMA
observations is not sufficient to add to the ongoing debate whether
these masers are associated with outflows or discs \citep[see][and
  references therein]{debuizer2009}.

\subsection{Data Modelling}
\label{sub:data_modelling}

\subsubsection{Temperature and density profile modelling of continuum data}
\label{subsub:cont_modelling}

\begin{figure*}
\begin{center}
 \includegraphics[width=12cm, angle=-90, trim=0 0 -5 0]{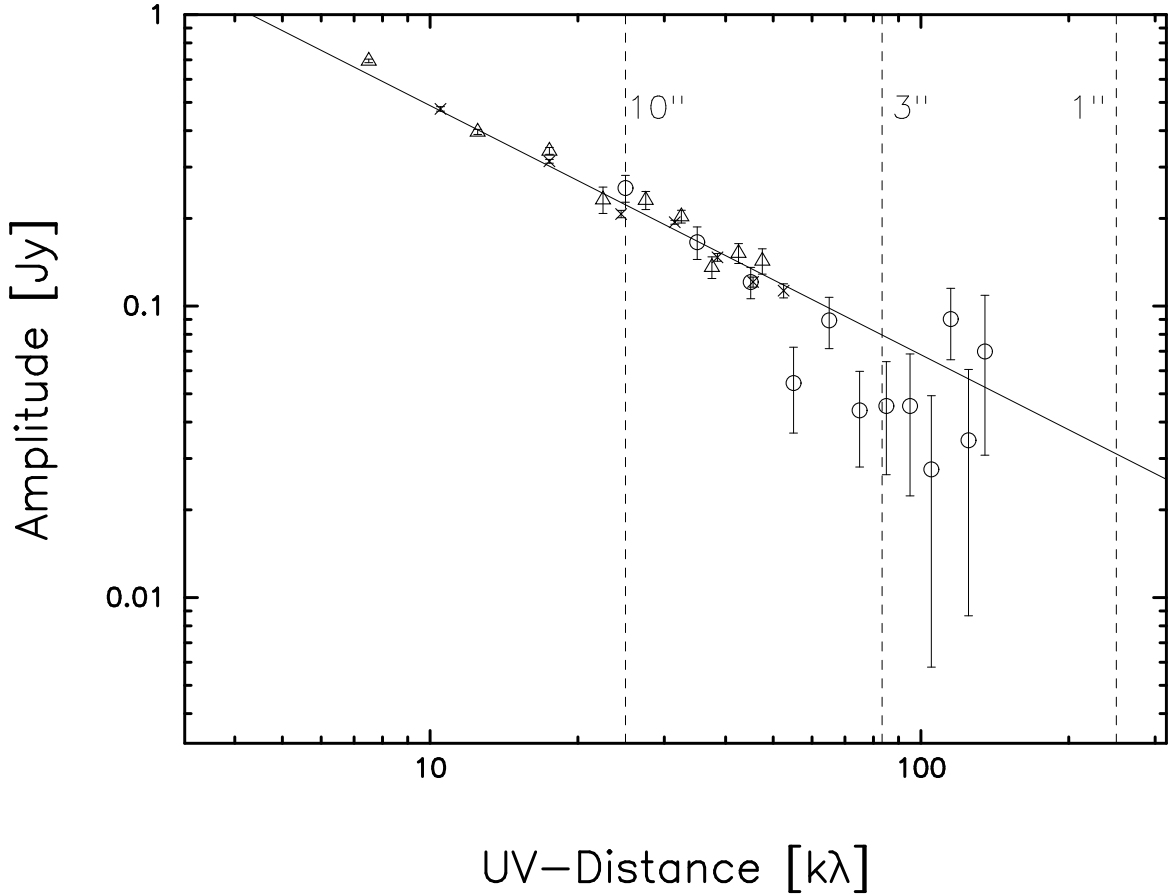}\\
\end{center}
\caption{UV-distance vs visibility amplitude for the 1mm continuum
  emission observed with the SMA. Visibilities from the compact,
  subcompact and extended array configurations are shown as triangles,
  squares and circles respectively. The error bars show the
  uncertainty calculated from the variation in amplitude for the
  visibilities binned for each point. The solid line shows the
  weighted least-squares fit to the data -- assuming T$ \propto
  r^{-0.33}$ as expected for an internally heated source (see
  $\S$~\ref{subsub:cont_modelling}) the best-fit density profile is
  given by $\rho \propto r^{-1.8\pm0.2}$. The dashed lines show the
  corresponding angular scale at 10$\arcsec$ (left), 3$\arcsec$
  (center) and 1$\arcsec$ (right). While the assumption of spherical
  symmetry appears a good fit to the data at large spatial scales, it
  is clear this assumption breaks down at $\lsim3\arcsec$, where the
  extended configuration observations resolve MM1, MM2 and MM3.}
\label{fig:g8.68_uvdistamp}
\end{figure*}

Following \citet{zhang2009} \citep[see also][]{takahashi2009}, we used
the measured visibility amplitude of the SMA continuum data as a
function of $uv$-distance to determine the spatial density profile of
G8.68$-$0.37. Assuming the density, $\rho$, and dust temperature,
T$_{\rm dust}$, scale as a power-law with radius (i.e. T$_{\rm
  dust}(r) \propto r^{-\kappa_T}$ and $\rho(r) \propto
r^{-\kappa_\rho}$) and the dust is centrally heated, then the flux
density from dust emission, $F$, is given by $F \propto \int \rho$
T$_{\rm dust} dS$, where S is the length along the line of sight. For
$\kappa_\rho + \kappa_T >1$, in the image domain the flux density
scales as $F \propto r^{-(\kappa_\rho + \kappa_T-1)}$, while in the
$uv$ plane this corresponds to $A \propto S_{uv}^{(\kappa_\rho +
  \kappa_T -3)}$, where $A$ is the visibility amplitude and $S_{uv}$
is the $uv$-distance. Figure~\ref{fig:g8.68_uvdistamp} shows the
visibility amplitude as a function of $uv$-distance for the subcompact
(triangles), compact (crosses) and extended (circles) array
configurations, scaled by the observing frequency. The error bars show
the uncertainty calculated from the variation in amplitude for the
visibilities binned for each point. Assuming a temperature exponent
$\kappa_T=0.33$ (appropriate for the dust being centrally heated), the
weighted least-squares fit to the data (shown as a line in
Figure~\ref{fig:g8.68_uvdistamp}) reveals $\kappa_\rho=1.8\pm0.2$. The
intercept of the fit, corresponding to the total flux, agrees well
with the value of 3.4\,Jy reported in \citet{hill2005}.

\subsubsection{Radiative transfer modelling of $\nhthree$ data}
\label{subsub:nh3_modelling}

To determine a second, independent measurement of the spatial
temperature and density profile we used radiative transfer modelling
to fit the \citet{L07A} $\nhone$, (4,4) and (5,5) data. The radiative
transfer code, MOLLIE\footnote{See \citet{keto1990} for a description
  of the MOLLIE code, \citet{keto2004} for a description of the
  line-fitting and \citet{carolan2009} and \citet{keto_zhang2010} for
  recent examples of work using the code}, can deal with arbitrary 3D
geometries, but based on (i) the morphology of the single-dish sub-mm
continuum emission, and (ii) analysis of the density profile in
$\S$\ref{subsub:cont_modelling}, a spherically symmetric model was
chosen as a good approximation, at least down to spatial scales of
$\sim$0.05\,pc probed by the extended configuration SMA
observations. A model radius of 0.58\,pc was chosen based on the
extent of the single-dish sub-mm continuum emission. The temperature,
T$_K$, and density, $\rho$, were again parameterized with radius, $r$,
as power laws: $\rho(r) = \rho_{1/2} \; r ^{-\kappa_\rho}$ and T$_K(r)
= T_{1/2}\; r ^{-\kappa_T}$, where $\rho_{1/2}$ and $T_{1/2}$ are the
density and temperature at the half-radius, 0.29\,pc.  An additional
constant velocity component, $\Delta V_{\rm NT}$, was included to
model non-thermal support (e.g. from turbulence), resulting in a five
parameter model for the cloud. The $\nhthree$ to H$_2$ abundance was
fixed at 5$\times$10$^{-9}$ \citep[derived from the observations of
  G8.68$-$0.37 in][]{pillai2007}. This is lower than typical
abundances found in young massive star formation regions but
increasing the abundance to several 10$^{-8}$ does not affect the
general results of the paper.

Models were constructed over a range of values for the five parameters
(see Table~\ref{tab:model_pars}).  Radiative transfer modelling was
then used to generate synthetic data cubes for the $\nhone$, (4,4) and
(5,5) emission. These were then convolved with 2D Gaussian profiles at
a spatial scale corresponding to the resolution (synthesized beam) of
the \citet{L07A} observations. The synthetic spectra at each
transition and position were fit to the observed spectra at three
positions and reduced-$\chi^2$ values returned for the
goodness-of-fit. The first position was located at the peak of the
sub-mm continuum emission ($\alpha_{J2000}=$18:06:23.47,
$\delta_{J2000}=-$21:37:7.6), then in radial steps of
$\sim$10$\arcsec$ (the synthesized beam size) at
$\alpha_{J2000}=$18:06:23.83, $\delta_{J2000}=-$21:37:15.6 and
$\alpha_{J2000}=$18:06:24.33, $\delta_{J2000}=-$21:37:21.6. While the
$\nhthree$ integrated intensity emission shows a similar extent to the
single-dish continuum emission, the $\nhthree$ morphology is extended
along the outflow axis. As $\nhthree$ is known to be affected by
outflow interactions \citep{zhang2007nh3outflow}, it is possible the
$\nhthree$ emission may be affected along this axis. Therefore, the
radial direction was chosen to be that perpendicular to the molecular
outflow (see $\S$~\ref{sub:outflow} and
Figures~\ref{fig:g8.68_rgb}~and~\ref{fig:g8.68_12co_chmap}) to
minimize any potential contamination.

\begin{figure*}
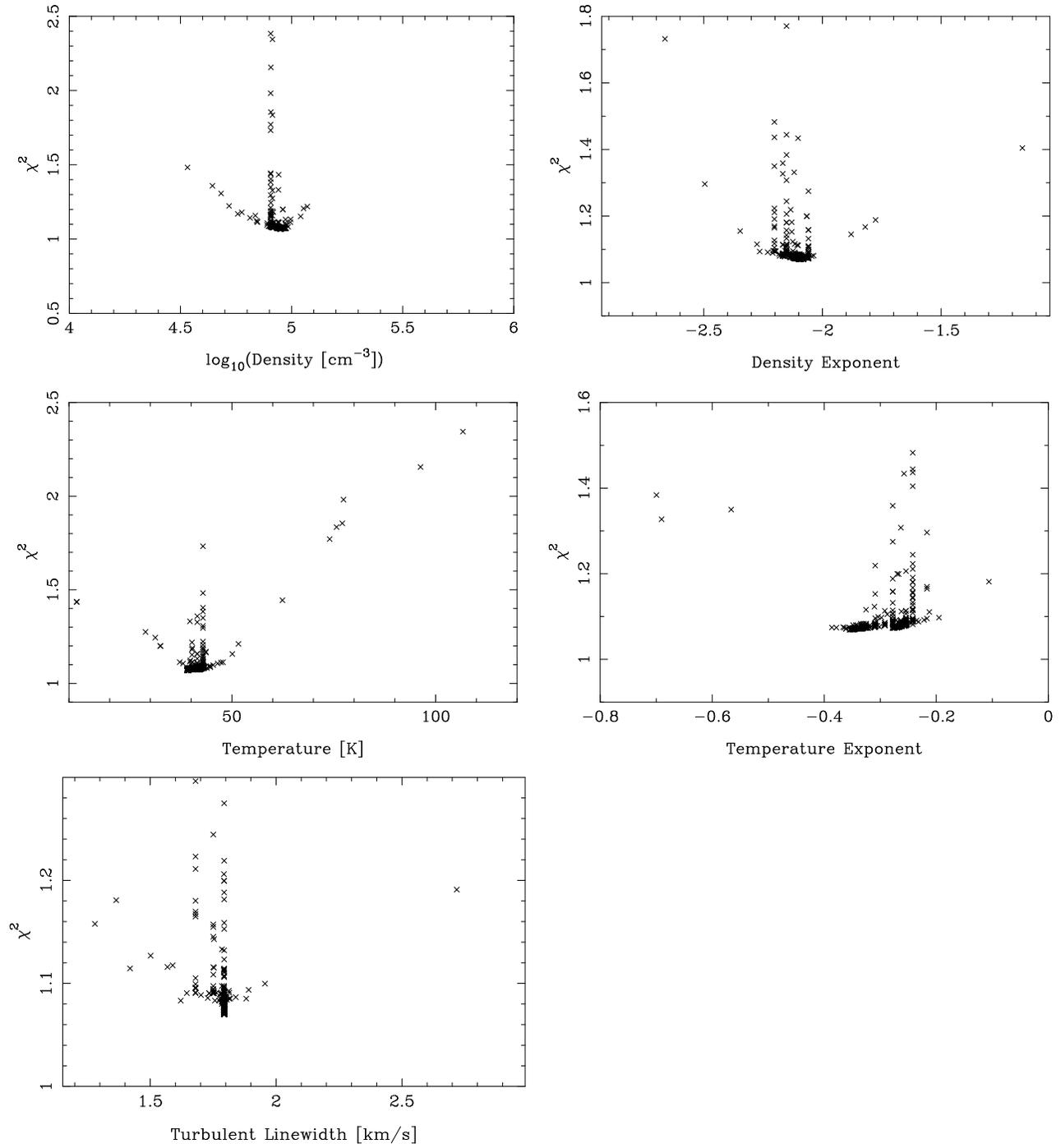

\begin{center}
  \begin{tabular}{cc}
    \includegraphics[width=6cm, angle=-90, trim=0 0 -5 0]{density_red_chisq.ps}&
    \includegraphics[width=6cm, angle=-90, trim=0 0 -5 0]{dens_exp_red_chisq.ps}\\

    \includegraphics[width=6cm, angle=-90, trim=0 0 -5 0]{temp_red_chisq.ps}&
    \includegraphics[width=6cm, angle=-90, trim=0 0 -5 0]{temp_exp_red_chisq.ps}\\

    \includegraphics[width=6cm, angle=-90, trim=0 0 -5 0]{turb_width_red_chisq.ps}\\
 
 \end{tabular}
\end{center}
\caption{Reduced-$\chi^2$ values from the simulated annealing process
  to find the best-fit model to the \citet{L07A} data (see
  $\S$~\ref{subsub:nh3_modelling}). Each of the crosses shows the
  physical parameters (density, density exponent, temperature,
  temperature exponent and turbulent linewidth) from one of the 10,000
  models, and the resulting reduced-$\chi^2$ value for that particular
  model.}
\label{fig:g8.68_rt_chisq}
\end{figure*}

Simulated annealing with 10,000 models was used to search through the
5D parameter space to minimize $\chi^2$ and find the best-fit
model. Figure~\ref{fig:g8.68_rt_chisq} shows the reduced-$\chi^2$
values for the range of parameter space covered by one such run
through the simulated annealing process. This method is inherently
robust against becoming trapped in local, rather than global minima in
parameter space. However, to determine the robustness of the best-fit
model we ran the fitting 20 times with widely separated initial start
values and increments. The results are similar to those in
Figure~\ref{fig:g8.68_rt_chisq}. Table~\ref{tab:model_pars} lists the
range of parameter space covered in the fitting process and the
resulting best-fit values. Figure~\ref{fig:g8.68_rt_fit} shows the
\citet{L07A} $\nhone$, (4,4) and (5,5) spectra from each of the three
positions overlayed with the synthetic spectra at the same position
from the best-fit model.

\subsubsection{$\methylcyanide$ modelling}
\label{subsub:ch3ch_modelling}

\begin{figure*}
\begin{center}
\begin{tabular}{c}
 \includegraphics[width=12cm, angle=0]{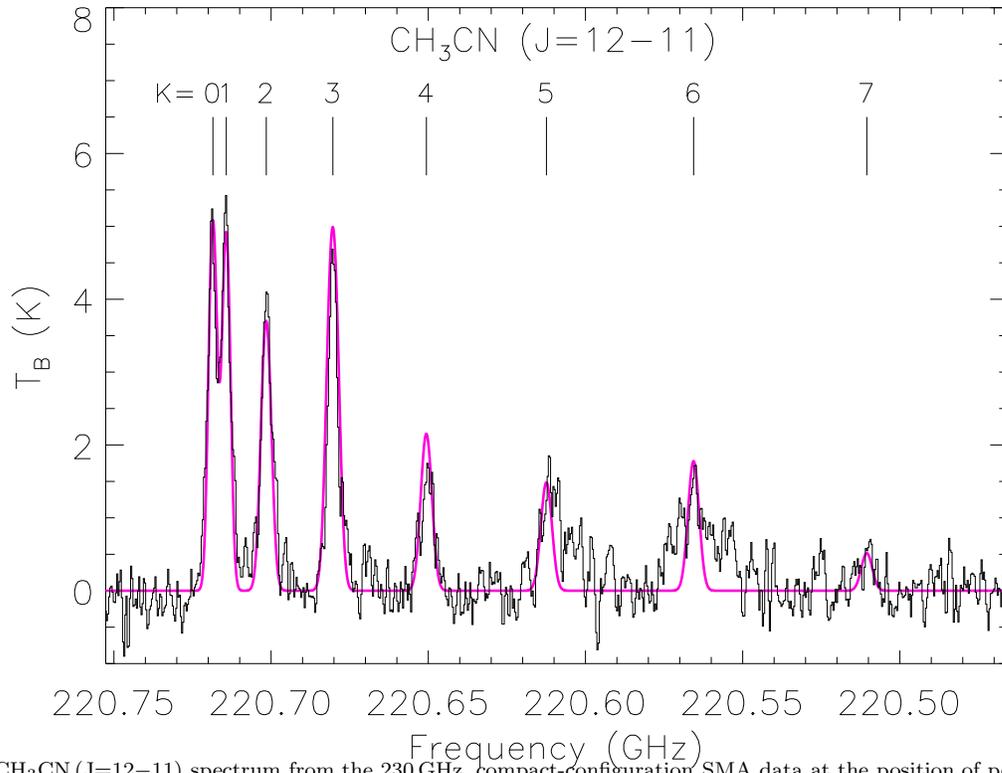}\\
\\
\end{tabular}
\end{center}
\vspace{3mm}
\caption{\methylcyanide\,(J$=$12$-$11) spectrum from the 230\,GHz,
  compact-configuration SMA data at the position of peak continuum
  emission [black] overlayed with the single-component LTE model fit
  as outlined in $\S$~\ref{subsub:ch3ch_modelling}. The K-components
  are labelled above the spectrum.}
\label{fig:g8.68_ch3cn_fit}
\end{figure*}

Emission from the \methylcyanide\, J$=$12$-$11 transition
[$\sim$220.7\,GHz] is detected in the K~$=$~0~to~7
components\footnote{Upper energy levels for K~$=$~0~to~7 components
  are 69, 76, 97, 133, 183, 246, 325, 417\,K, respectively} in the
230\,GHz compact-configuration observations. The emission is
unresolved in the 3\arcsec$\times$2\arcsec\, beam and at the peak of
the mm continuum emission, encompassing the location of MM1 to
MM3. Physical parameters of the gas at this spatial scale were
estimated by $\chi^2$ minimization fitting of the spectrum, solving
for the optical depth assuming LTE conditions. With a collisional rate
of 10$^{-8}$\,cm$^3$\,s$^{-1}$ \citep{green1986} and Einstein A
coefficient of 10$^{-4}$\,s$^{-1}$ the transition has a nominal
critical density of 10$^4$\,$\cm3$ so the excitation should be
collisionally dominated in densities typical of massive star forming
cores.  The observed spectrum and resultant fit are shown in
Figure~\ref{fig:g8.68_ch3cn_fit}. While the single-component model
clearly fits the data well, the increasing linewidth and slight excess
of emission in the high-K vs low-K components means the emission
probably has more than one component, and the gas potentially has a
temperature and density gradient. However, the signal-to-noise of the
higher K-components is not sufficient to constrain more detailed
modelling. The best-fit temperature including all K-components is
200\,K which was robust to the 10\% level under variation of the
initial fitting estimates and number of free parameters.  Only fitting
the K$=$0 to 4 components reduces the temperature to 100\,K but gives
a much poorer result for the higher K components. The best-fit
\methylcyanide\, column density is 10$^{16}$cm$^{-2}$. Given typical
abundances \citep{wilner1994, zhang1998, hatchell1998, chen2006,
  zhang2007, leurini2007}, the total H$_2$ column density at this size
scale is at least 10$^{24}$cm$^{-2}$ and possibly significantly
larger. For the above models, the filling factor implies a size of
0.8$\arcsec$ for the $\methylcyanide$ emission region.

\subsubsection{Model Consistency}
\label{subsub:model_consistency}

We now compare the results of the modelling outlined in
$\S$~\ref{subsub:cont_modelling}, \ref{subsub:nh3_modelling} and
\ref{subsub:ch3ch_modelling} to check for consistency.  At larger
spatial scales, the results of the continuum and $\nhthree$ modelling
agree well, suggesting the simple spherical model with power-law
temperature and density gradient is a good approximation at this
spatial scale. However, the fit to the SMA continuum data in
Figure~\ref{fig:g8.68_uvdistamp} shows the assumption of spherical
symmetry breaks down at angular scales below a few arcseconds --
corresponding to the size at which the extended configuration resolves
the emission into the three cores, MM1 to MM3. In the subsequent
discussion section we refer to this radius of $\sim$0.05\,pc as the
`minimum reliable radius' for the power-law density and temperature
approximation.

Given the poorer resolution ($\geq$8\arcsec), the observed $\nhthree$
emission will be dominated by gas at much larger spatial
scales. However, extrapolating the temperature and density gradient
derived from the $\nhthree$ data down to 0.8$\arcsec$ (the best-fit
size of the $\methylcyanide$ emission) predicts a temperature of
100\,K and density of 2$\times$10$^7$\cm3, which matches well with
properties derived from the $\methylcyanide$ data when only fitting
the lower K-components. The excess emission and linewidth in the
higher $\methylcyanide$ K-components points to the density and
temperature gradient continuing at even smaller spatial scales. In
summary, modelling of the three different data sets produces a
coherent picture of the density and temperature profile of
G8.68$-$0.37.

\section{Masses of MM1 to MM3}
\label{sec:mass_mm1_to_mm3}

We calculate the mass of MM1 to MM3 using the temperatures derived
from the $\nhthree$ and $\methylcyanide$ data -- $\sim$100\,K at the
radius encompassing all three components and likely increasing to
$\sim$200\,K at smaller radii. Without independent temperature
measurements of each component, the plausible mass range of MM1 to
MM3, assuming standard dust properties and following
\citet{kauffmann2008}, is then 7$-$14\,M$_\odot$, 5$-$9\,M$_\odot$ and
2$-$5\,M$_\odot$, respectively. This is a small fraction of the total
mass of 1500\,M$_\odot$ derived from the single-dish observations
\citep{hill2010}.

\section{Discussion}
\label{sec:discussion}

We now return to the original focus of the paper to investigate the
case for thermal feedback affecting fragmentation in this young
massive protocluster.

We consider a simplistic scenario where, at $t~=~t_0$, the region
initially had conditions similar to those of IRDCs: $T_0 \sim 10-20K$
and $n_0\sim10^4$\,\cm3. A first generation of stars formed from this
gas and began to heat up the surrounding environment. Some time later,
at $t~=~t_{\rm frag}$, the more massive cores we observe with the SMA
formed from thermal fragmentation of the gas. The gas temperature and
density at t$_{\rm frag}$ was T$_{\rm frag}$ \& n$_{\rm
  frag}$. Assuming purely thermal fragmentation, this gas temperature
and density set the mass (M$_{\rm frag}$) and separation
($\lambda_{\rm frag}$) of the resulting fragments. Some further time
later, at $t~=~t_{\rm now}$, we observed the region and have
determined the present-day temperature [T$_{\rm now}$(r)] and density
[$\rho_{\rm now}$(r)] as a function of radius from $r=0.05$ to
0.58\,pc and the current mass (M$_{\rm now}$) and separation
($\lambda_{\rm now}$) of cores MM1 to MM3 resolved on smaller scales.

In the following sections we use these observational results to infer
the possible conditions at $t_{\rm frag}$ (i.e. T$_{\rm frag}$,
$\rho_{\rm frag}$, $\lambda_{\rm frag}$ and M$_{\rm
  frag}$\footnote{$\lambda_{\rm frag}$ and M$_{\rm frag}$ are
  equivalent to the Jeans length and Jeans mass,
  respectively.}). Given these inferred conditions, we then
investigate the plausibility of different sources of feedback raising
the gas temperature to the required T$_{\rm frag}$. In this way we aim
to assess whether thermal feedback may have affected fragmentation.

In $\S$~\ref{sub:static} we start with the assumption that the
present-day density profile and core mass/separations are similar to
those at the time of fragmentation -- so $\rho_{\rm now} = \rho_{\rm
  frag}$, $\lambda_{\rm now} = \lambda_{\rm frag}$ and M$_{\rm
  now}=$M$_{\rm frag}$. In $\S$~\ref{sub:dynamic} we relax these
assumptions.

\subsection{Thermal fragmentation assuming $\rho_{\rm now} = \rho_{\rm frag}$, $\lambda_{\rm now} = \lambda_{\rm frag}$  and $M_{\rm now}=M_{\rm frag}$}
\label{sub:static}

\subsubsection{Inferring T$_{\rm frag}$}
\label{subsub:static_tfrag}


Assuming MM1 to MM3 have not moved since fragmentation, their relative
positions in Figure~\ref{fig:g8.68_rgb} then correspond to their
fragmentation separations, $\lambda_{\rm frag}$, or Jeans length,
projected on to a 2D plane. The measurement uncertainty in the
projected separations is small -- of order a few percent for the
significance of the MM1 to MM3 detections \citep[see e.g.][Eq
  14-5]{fomalont1999}, or calculated from the r.m.s phase noise,
$\Delta\phi$, and synthesized beam, $\theta_B$ (positional uncertainty
$\sim$ $\Delta\phi \,\, \theta_B/2\pi$). The major systematic
uncertainty is the distance to the region, which we estimate to be
$\sim$20\%. As the measured separations are in projection, they
underestimate the real separations. We try to quantify this
underestimation through geometrical arguments. Making the reasonable
assumption that MM1 to MM3 lie within the volume encompassed by
$r<0.05$\,pc (as opposed to lying at much larger radii and only seen
in projection towards the center), the largest underestimate will
occur where a given pair lie at opposite sides (ie front vs back) of
the $r<0.05$\,pc volume. Assuming MM1 lies near the center, the
maximum separation is then 0.05\,pc. The projected separations then
underestimate the real separations by $\leq$40\%. Any underestimate in
separation will be reflected in an underestimate of T$_{\rm frag}$.

Completeness is another potential observational bias in determining
$\lambda_{\rm frag}$. If we are missing lower-mass fragments which
fall below the detection limit, for example, the distance from MM1 to
MM3 to the nearest fragment may be lower than the measured projected
separations. However, from the sensitivity of the SMA observations
($\sigma_{\rm RMS} \sim 1-2$\,mJy), any isolated fragments above
$\sim$1\,M$_\odot$ should be detected by their dust continuum
emission. This 1\,M$_\odot$ limit is equal to the global Jean's mass
for the physical conditions typical of cold IRDCs prior to the
formation of any stars. A more in-depth discussion of the potential
existence of a population of pre-existing but undetected lower mass
stars is given in $\S$~\ref{subsub:heating_by_lower_mass_stars}.


A final concern in determining $\lambda_{\rm frag}$ might be that MM1
to MM3 are comprised of multiple, unresolved dust continuum peaks. For
example, $\sim$0.2$\arcsec$ angular resolution observations towards
similar regions at a similar distance and evolutionary stage resolve
multiple (proto)stars with linear separations down to 1700AU
\citep{longmore2006}.  However, such systems would correspond to
multiple stars forming within a single fragment, rather than
reflecting the spatial distribution of fragments at t$_{\rm frag}$.

The uncertainties in the mass of MM1 to MM3, and hence M$_{\rm frag}$,
are far larger than those of $\lambda_{\rm frag}$. In addition to the
large uncertainty introduced by the lack of independent temperature
measurements for MM1 to MM3 (see $\S$~\ref{sec:mass_mm1_to_mm3}),
there are major systematic uncertainties in deriving gas masses from
thermal dust emission -- e.g. dust properties, distance\footnote{The
  uncertainty in mass is proportional the uncertainty in distance,
  $\Delta D$, squared, rather than linearly proportional to $\Delta D$
  as is the case for the uncertainty in the measured separations.},
gas-to-dust ratio etc.

The density distribution is constrained at the larger spatial scales
by the $\nhthree$ modelling ($\S$~\ref{subsub:nh3_modelling}) and at
arcsecond scales by the $\methylcyanide$ modelling
($\S$~\ref{subsub:ch3ch_modelling}). As discussed in
$\S$~\ref{subsub:model_consistency}, these independently predict a
density of $\sim$2$\times$10$^7 \cm3$ at the size scales of MM1 to
MM3. This provides an upper limit to $\rho_{\rm frag}$ as the rest of
the volume containing MM1 to MM3 is likely to be at a lower
density. As a more realistic estimate of $\rho_{\rm frag}$, the
density profile derived in $\S$~\ref{sub:data_modelling} predicts an
average density of 8$\times$10$^6 \cm3$ within a volume of
$r<0.05$\,pc (the radius encompassing MM1 to MM3).

Figure~\ref{fig:g8.68_tfrag} encapsulates the above uncertainties and
illustrates how these affect the inferred fragmentation
temperature. The solid lines show the expected T$_{\rm frag}$ as a
function of density for Jeans lengths of 0.05 and 0.03\,pc,
corresponding to upper and lower limit estimates of the MM1 to MM3
separations. The dotted lines show the expected T$_{\rm frag}$ as a
function of density for Jeans masses of 3 and 10\,M$_\odot$,
corresponding to reasonable estimates of the MM1 to MM3 masses. This
shows that for the predicted densities, the fragmentation temperature
must be at least 100\,K.

We now consider the plausibility of different sources of feedback
raising the gas temperature to this required T$_{\rm frag}$.

\subsubsection{The case for heating by pre-existing, unobserved, embedded lower-mass stars}
\label{subsub:heating_by_lower_mass_stars}

From the sensitivity of the SMA observations ($\sigma_{\rm RMS} \sim
1-2$\,mJy), any isolated dust continuum fragments above
$\sim$1\,M$_\odot$ should be detected, but the SMA would not detect
unembedded stars. The Spitzer observations
(e.g. Figure~\ref{fig:g8.68_rgb}) would detect stars, but not if they
were very deeply embedded. Due to spatial filtering, interferometers
are only sensitive to density contrasts, so it is possible the
observations are missing a population of uniformly-distributed,
embedded lower mass stars.

To assess the feasibility that pre-existing and unobserved stars are
responsible for heating G8.68$-$0.37, we investigate what stellar
population would be required to generate the required 100\,K thermal
fragmentation temperature. We take the simple approach of calculating
the effect of heating by individual stars of different mass and
luminosity to find the radius out to which they will heat the gas to
100\,K. Given this radius, we then determine the minimum number of
these stars uniformly spaced within the total spherical volume of
radius 0.05\,pc necessary to raise the temperature to 100\,K.

\begin{figure*}
\begin{center}
\begin{tabular}{c}
\includegraphics[width=12cm, angle=-90]{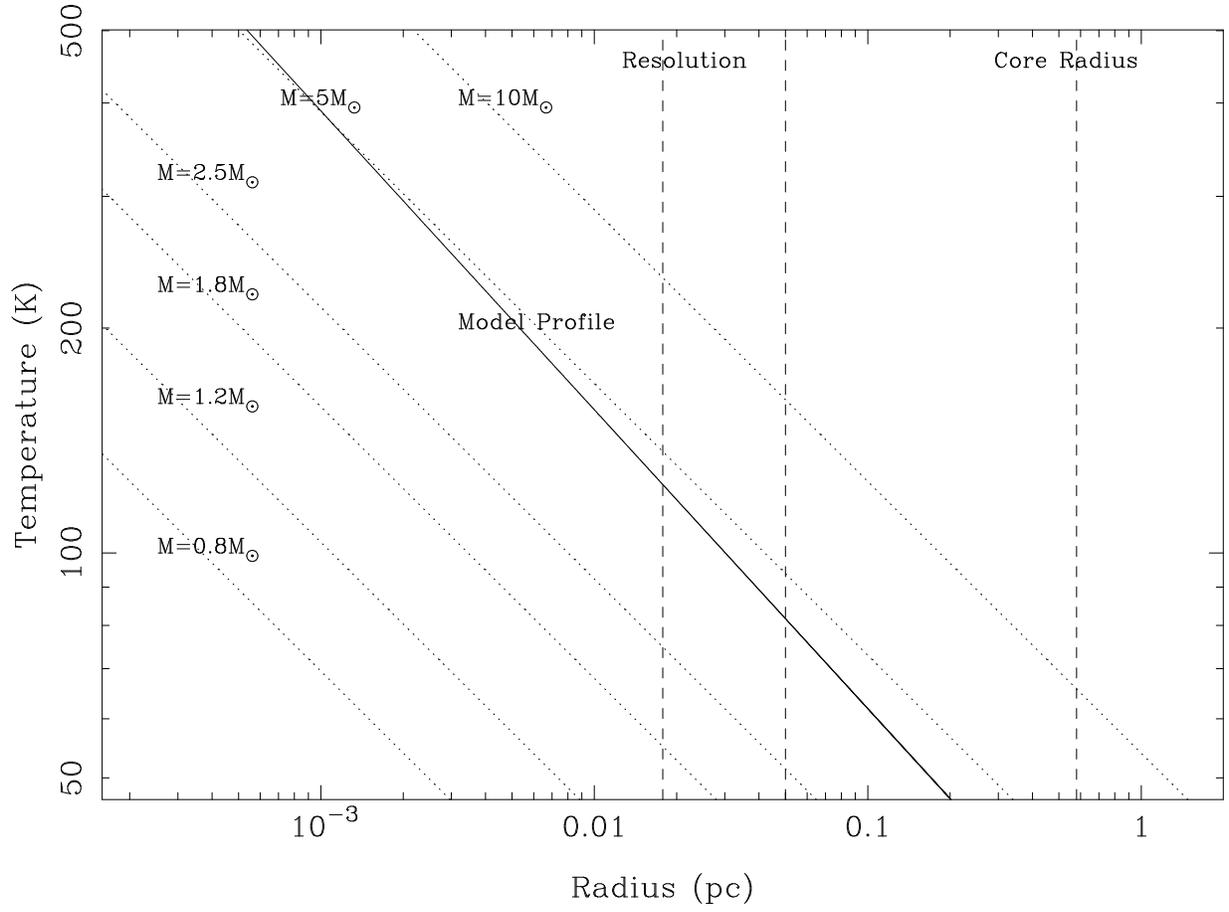} \\
\\
\end{tabular}
\end{center}
\vspace{3mm}
\caption{Temperature as a function of distance from stars of various
  mass.  The solid line shows the temperature of the best-fit model
  determined in $\S$~\ref{sub:data_modelling} (T$ \propto r^{-0.35}$
  and $\rho \propto r^{-2.08}$). The right and central vertical dashed
  lines show the size-scale over which this profile is reliable --
  from the core radius down to 0.05\,pc (where the mm-continuum
  emission is resolved into multiple components). The left vertical
  dashed line gives the highest resolution of the SMA
  observations. The dotted lines show the analytical relationship
  between dust temperature and radius, assuming the dust is being
  heated and is in radiative equilibrium with a central star
  \citep[][Eq. 7.37]{lamers_cassinelli1999}. To ensure an upper-limit
  to the stellar temperature, the stars are assumed to have reached
  the main-sequence \citep[][Table
    15.14]{allen_ast_quant2000}. Temperature profiles are shown for
  stars of mass~$=0.8, 1.2, 1.8, 2.5, 5$ and $10$\,M$_\odot$. }
\label{fig:g8.68_tdust_rdust_mass}
\end{figure*}

Assuming the stars and surrounding dust are in radiative
equilibrium\footnote{i.e. each radius is considered as a geometrically
  thin and optically-thick shell with no opacity between it and the
  star.} and the opacity can be approximated by a power-law of
wavelength proportional to $\lambda^{-p}$, a star with temperature,
$T_\star$, and radius, $R_\star$, will raise the dust to temperature,
$T_d$ at a radius, $r_d$ ($r_d>>R_\star$), given by $r_d \simeq
(R_\star/2)(T_d/T_\star)^{-(4+p)/2}$
\citep[][Eq. 7.37]{lamers_cassinelli1999}. We adopt $p = 1.5$ based on
the opacities determined from the MRN \citep{mrn1977} grain-size
distribution between wavelengths of 0.1$\mu$m and 1mm
\citep{krugel_siebenmorgen1994}. To investigate the case of maximal
heating, the central stars are assumed to have reached the main
sequence. This provides an upper limit to the stellar effective
temperature because protostars on their Hayashi tracks are generally
cooler \citep[][Table
  15.14]{allen_ast_quant2000}. Figure~\ref{fig:g8.68_tdust_rdust_mass}
shows the resulting dust temperature as a function of radius from
stars of mass~$=0.8, 1.2, 1.8, 2.5, 5$ and $10$\,M$_\odot$, overlayed
with the best-fit model profile determined in
$\S$~\ref{sub:data_modelling}. Main-sequence stars of 0.8, 1.2 and
$1.8$\,M$_\odot$ will heat dust to $\geq$100\,K at radii $\leq$
3.5$\times$10$^{-4}$, 1.0$\times$10$^{-3}$ and
3.5$\times$10$^{-3}$\,pc, respectively. To heat the volume of radius
0.05\,pc to 100\,K in this way therefore requires uniformly
distributing these stars by the above separations throughout the
volume.  The minimum number of stars this implies (2.9$\times$10$^6$,
1.25$\times$10$^5$ and 2.9$\times$10$^3$) within a radius of 0.05\,pc
leads to total stellar masses far in excess of the total mass of gas
available. We therefore conclude the number of stars required to heat
the gas is unfeasibly large.

\begin{figure*}
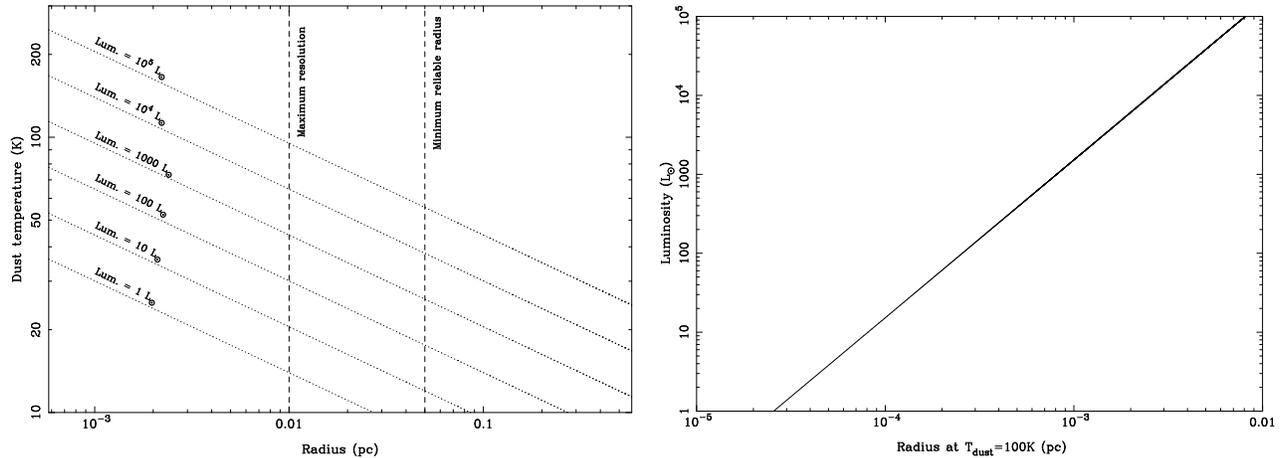

\begin{center}
\begin{tabular}{cc}
\includegraphics[width=6cm, angle=-90]{g8_68_sk.ps} &
\includegraphics[width=6cm, angle=-90]{g8_68_sk_r_at_t100.ps} \\
\\
\end{tabular}
\end{center}
\vspace{3mm}
\caption{[Left] Dust temperature as a function of radius for central
  heating sources of different luminosity.  The right and left
  vertical dashed lines show the radius where the mm-continuum
  emission is resolved into multiple components, and the highest
  resolution of the SMA observations, respectively.  The dotted lines
  show the analytical relationship between dust temperature and
  radius, assuming the dust is being heated and is in radiative
  equilibrium with a central powering source of the listed luminosity
  \citep[][Eq 9]{scoville_kwan1976}. [Right] Using the same analytical
  description in \citet{scoville_kwan1976}, the solid line shows the
  source luminosity required to heat the gas to $\geq$100\,K at the
  distance given on the x-axis.}
\label{fig:g8.68_tdust_rdust_lum}
\end{figure*}

At the earliest evolutionary stages, the luminosity from low-mass
stars can be dominated by accretion rather than nuclear burning. We
now consider the affect of heating due to accretion luminosity. As
most of the infalling particles' potential energy will be released at
radii close to the protostar, the luminosity can be approximated as
arising from a point-source. Following similar assumptions about the
dust properties etc. as before, the dust temperature as a function of
radius can be solved analytically \citep[e.g.][Eq
  11]{scoville_kwan1976,garay_lizano1999}. The left-hand panel of
Figure~\ref{fig:g8.68_tdust_rdust_lum} shows the dust temperature as a
function of radius for central heating sources with luminosities from
1$-$10$^5$\,L$_\odot$. This range was chosen to include the expected
luminosity from accreting low-mass stars (which are dominated by
accretion-luminosity) through to O-stars dominated by stellar
luminosity.  The right-hand panel of
Figure~\ref{fig:g8.68_tdust_rdust_lum} shows the radius at which
central heating sources of a given luminosity will heat the
surrounding gas up to $\geq$100\,K. Following the same line of
argument as before, we can investigate how many sources are required
to heat the 100\,M$_\odot$ volume of radius 0.05\,pc (the mass
determined from the modelling that lies within the radius encompassing
MM1 to MM3) to 100\,K as observed in
G8.68$-$0.37. \citet{froebrich2005} show that isolated Class~0 cores
(which as the youngest protostars should have the highest accretion
rates) typically have luminosities $\leq$10\,L$_\odot$. Again, an
unfeasible number of low mass stars are needed to heat the gas to the
required temperature, even allowing for the additional heating from
accretion luminosity. Assuming the higher infall rates in a massive
protocluster environment give rise to larger accretion luminosities
does not solve the problem either. The total luminosity required would
be larger than the present-day measured bolometric luminosity of
1.9$\times$10$^4$\,L$_\odot$.


Comparing the \citet{lamers_cassinelli1999} temperature profiles to
those of \citet{garay_lizano1999}\footnote{We compared the two
  analytic solutions by taking the known radii and effective
  temperatures of ZAMS stars of a given luminosity}, we note the
former are factors of a few higher. The \citet{garay_lizano1999}
formulation appears closer to observed profiles \citep[e.g.][and the
  gas temperatures derived from our own observations
  ($\S$~\ref{sub:data_modelling})]{keto_zhang2010} so we consider the
\citet{lamers_cassinelli1999} formulation
(e.g. Figure~\ref{fig:g8.68_tdust_rdust_mass}) as an upper limit to
the temperature.


The discrepancy between these two methods does not affect the
conclusion from the above analysis. A large population of low mass
stars would not be able to heat the gas to the 100\,K thermal
fragmentation temperature required to reproduce the observed
separations and masses of cores MM1 to MM3. This is consistent with
the models of \citet{offner2009} which show that low-mass stars have a
negligible affect on the overall cloud heating. Repeating the above
calculations but imposing more realistic limits -- that the total
stellar mass is not larger than the available gas mass and the total
luminosity is not larger than the measured bolometric luminosity --
gives an upper limit to the temperature provided by low-mass stars of
$\sim$40\,K.

\subsubsection{The case for heating by an early-B star}
\label{subsub:heating_by_5msun_star}

The calculations in the previous section show that low-mass stars
would not be able to heat the gas within a radius of 0.05\,pc to the
temperature of 100\,K required for the observed cores (MM1, MM2 and
MM3) to have formed by thermal fragmentation. The solid line
representing the best-fit model temperature profile in
Figure~\ref{fig:g8.68_tdust_rdust_mass} shows that this heating could
be attributed to a single early B-type star rather than a larger
number of low-mass stars. However, invoking an early B-type star as a
source of thermal feedback to raise the gas temperature and suppress
fragmentation (allowing cores as massive as MM1, MM2 and MM3 to form)
is unsatisfactory as it invokes a circular argument -- it does not
solve the problem of what initially raised the temperature to suppress
fragmentation into lower mass fragments which would have allowed the B
star to form in the first place.

\subsection{Thermal fragmentation relaxing the assumptions that $\rho_{\rm now} = \rho_{\rm frag}$, $\lambda_{\rm now} = \lambda_{\rm frag}$  and $M_{\rm now} = M_{\rm frag}$}
\label{sub:dynamic}

Having concluded it is not possible for thermal feedback to have
raised the gas temperature to $\geq$100\,K if $\rho_{\rm now} =
\rho_{\rm frag}$, $\lambda_{\rm now} = \lambda_{\rm frag}$ and $M_{\rm
  now} = M_{\rm frag}$, we consider how changes in the gas conditions
from $t_{\rm frag}$ to the present-day could have affected the
resulting fragments and look for observational predictions that these
changes would imply.

As shown in Figure~\ref{fig:g8.68_tfrag}, the required fragmentation
temperature would be lower than 100\,K if the density at the time of
fragmentation were lower than the present day -- ie if $\rho_{\rm now}
> \rho_{\rm frag}$. In $\S$~\ref{subsub:global_infall} we investigate
the observational evidence for the large-scale infall of gas that this
would imply. In $\S$~\ref{subsub:sf_on_the_move} we consider how a
lower density of gas at $t_{\rm frag}$ would have affected the
separations of MM1 to MM3 over time.

\subsubsection{The case for large-scale infall}
\label{subsub:global_infall}

As discussed in $\S$~\ref{sub:iras18032}, the molecular line profiles
in \citet{purcell2009} provide observational evidence for large-scale
infall of gas in G8.68$-$0.37. Following the calculations in
\citet{walsh2006}, the infall rate for G8.68$-$0.37 inferred from the
HCO$^+$ and HCN spectra is $\sim$10$^{-4}$\,M$_\odot$\,yr$^{-1}$. This
infall rate is significantly larger than typically measured towards
lower mass star-forming cores but comparable to observations of other
young high mass star formation regions
\citep{wu2003,peretto2006,peretto2007,walsh2006,keto_klassen2008,
  dewit2009}.

\begin{figure*}
\begin{center}
\begin{tabular}{c}
 \includegraphics[width=10cm, angle=-90]{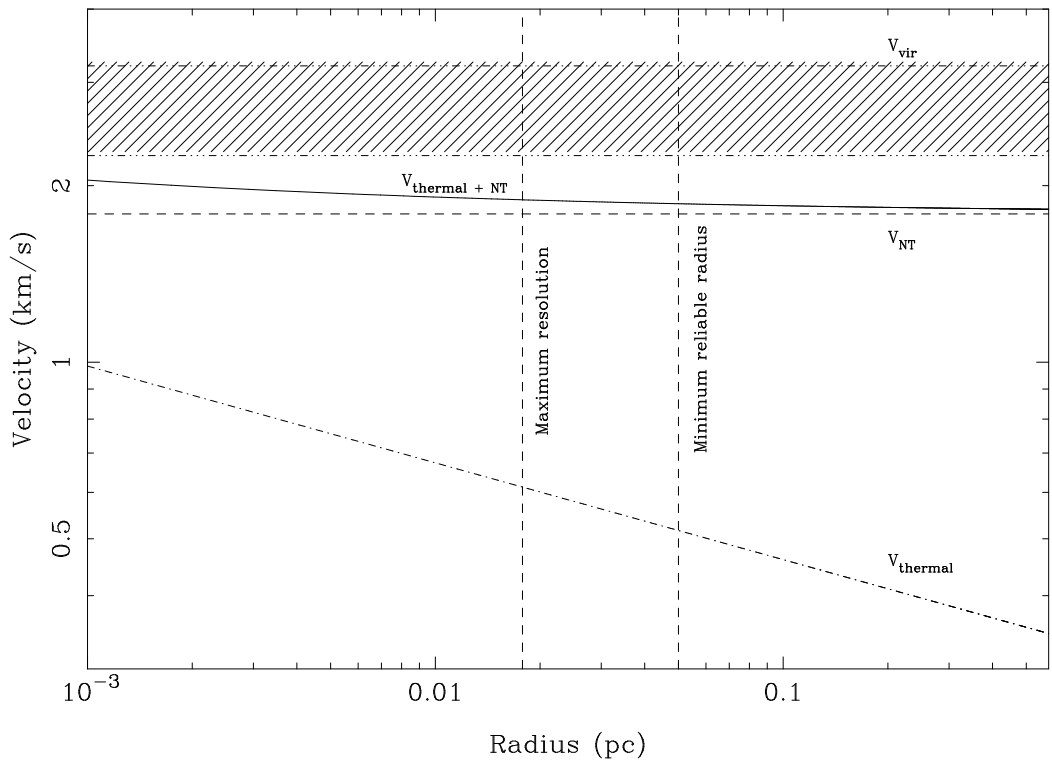}\\
\end{tabular}
\end{center}
\vspace{3mm}
\caption{Comparison of non-thermal and thermal support in
  G8.68$-$0.37. The horizontal dashed line, indicated as V$_{\rm NT}$,
  shows the non-thermal contribution to the line-width determined from
  the $\nhthree$ radiative transfer modelling in
  $\S$~\ref{subsub:nh3_modelling}. The dot-dash line, labelled $V_{\rm
    thermal}$, shows the expected thermal contribution to the
  $\nhthree$ line-width as a function of radius, determined from the
  temperature distribution of the best-fit model in
  $\S$~\ref{sub:data_modelling}. The solid line, $V_{\rm thermal +
    NT}$, shows V$_{\rm NT}$ and $V_{\rm thermal}$ added together in
  quadrature giving the expected total linewidth. Finally, the
  dash-dot-dot lines and enclosed hatched area gives the expected
  range in virial velocity, $V_{\rm virial} \equiv \left (
  \frac{GM}{R} \right )^{1/2}$ -- a measure of how much kinetic energy
  is required to balance the gravitational potential of a cloud. The
  right and left vertical dot-dashed line shows the minimum reliable
  radius for the assumed power-law density distribution and the
  maximum resolution of the SMA observations, respectively. This
  figure shows, i) as expected the linewidth is dominated by the
  non-thermal contribution, and ii) even the higher temperatures
  towards the center are insufficient to reach the required kinetic
  energy support for the cloud to be in equilibrium.  }
\label{fig:g8.68_vir_vel}
\end{figure*}

Are these infall motions suggested by the molecular line observations
consistent with the velocities expected in gravitational collapse? We
estimate the dynamical state of the region by comparing the
contribution from thermal and non-thermal support as a function of
radius to the virial velocity, $V_{\rm virial} \equiv \left
(\frac{GM}{R} \right )^{1/2}$ \citep[][Eq 3.20]{stahler_palla2005} --
a measure of how much kinetic energy is required to balance the
gravitational potential. To convert from the measured one-dimensional
velocity to the three-dimensional root mean square velocity we use
\citet{rohlfs_wilson2004}, Eq 12.72. The virial velocity in
Figure~\ref{fig:g8.68_vir_vel} is then calculated at each radius using
the enclosed mass (derived from the model parameters in
$\S$~\ref{sub:data_modelling}) inside that radius. The range of
uncertainty in virial velocity \citep[see e.g.][]{elmegreen1989} is
illustrated by the hatched area between the dash-dot-dot
lines. Figure~\ref{fig:g8.68_vir_vel} shows, i) as expected the
linewidth is dominated by the non-thermal contribution, and ii) even
the higher temperatures towards the center are insufficient to reach
the required kinetic energy support for the cloud to be in
equilibrium. Although undoubtedly oversimplifying the gas dynamical
state (both support from magnetic fields and the surface terms in the
virial equation are ignored, for example), we find that the inward
motions are generally consistent with gravitational contraction.

\subsubsection{Could the cores have been more widely separated in the past?}
\label{subsub:sf_on_the_move}

If the entire star forming cloud is globally contracting, then the
density of the gas from which MM1 to MM3 fragmented could have been
lower than measured today. The cores could initially have formed at
larger separations and moved closer over time to reach their current
locations. To test the feasibility of this scenario we can estimate
how far these cores may have moved over a given time period and
predict their resulting relative velocities. Assuming the
fragmentation conditions were similar to those of IRDCs
(10$^5$\,$\cm3$ \& $10-20$\,K), which are thought to be massive star
formation regions at the earliest evolutionary stages, the Jeans
length would have been $\sim$0.07\,pc. To reach their current
separations of $\sim$0.048\,pc, the cores would have had to move
$\sim$0.01\,pc towards each other.

\begin{figure*}
\begin{center}
\begin{tabular}{c}
 \includegraphics[width=12cm, angle=-90]{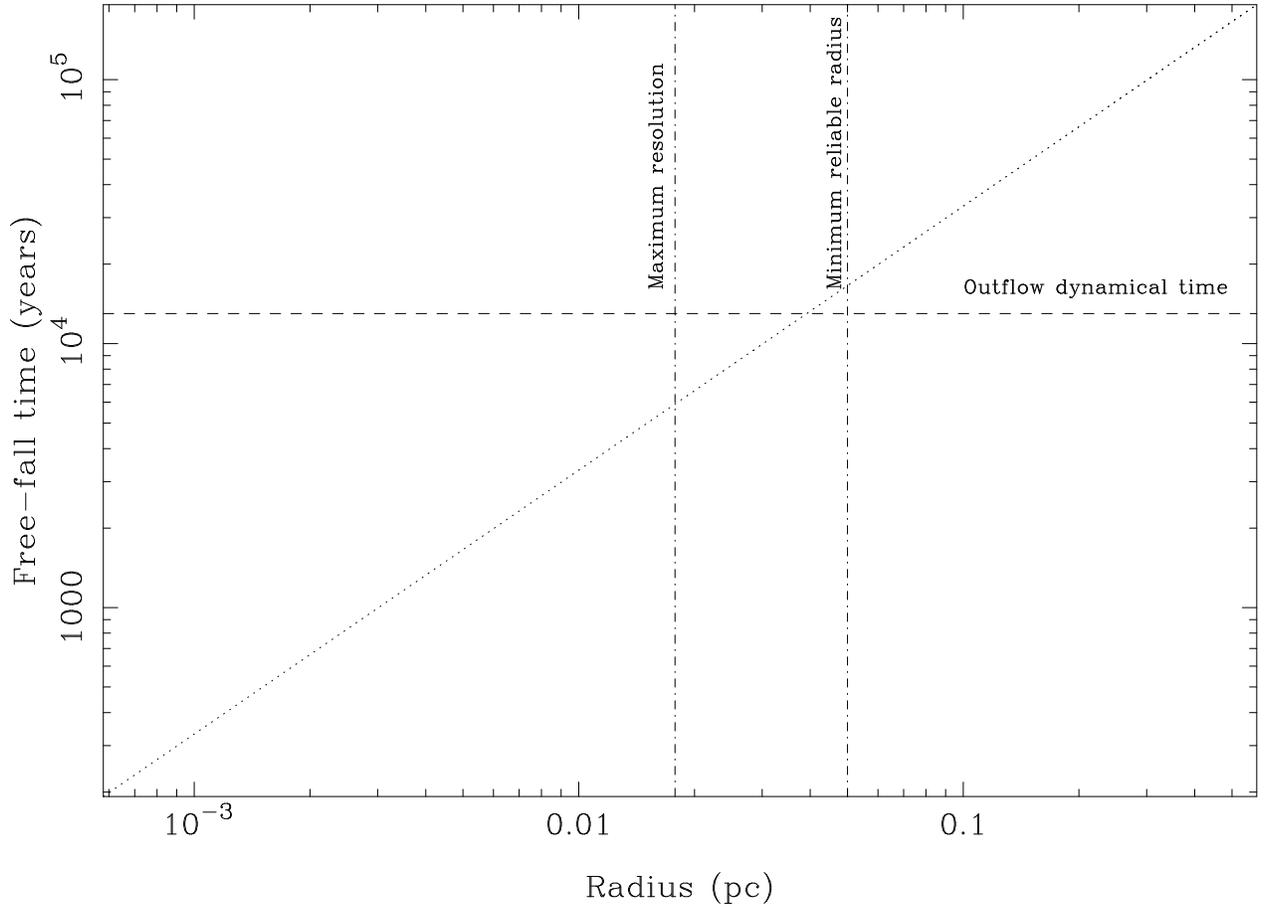}\\
\end{tabular}
\end{center}
\vspace{3mm}
\caption{The dotted line shows the free-fall time, $t_{ff} \equiv
  \left ( \frac{3\pi}{32 \, G \, \rho_{\rm av}} \right ) ^{1/2}$, as a
  function of radius where $\rho_{\rm av}$ is the average density
  enclosed within a given radius determined from the enclosed mass
  within the radius on the horizontal axis. The right and left
  vertical dot-dashed line shows the minimum reliable radius for the
  assumed power-law density distribution and the maximum resolution of
  the SMA observations, respectively. The dynamical time of the
  outflow determined in $\S$~\ref{sub:outflow} is shown as a dashed
  horizontal line. The free-fall time at the radius at which the core
  is observed to fragment is similar to the dynamical time of the
  outflow.}
\label{fig:g8.68_tff}
\end{figure*}

We use the free-fall time to derive a lower limit on how long the
cores may have been moving since they formed.
Figure~\ref{fig:g8.68_tff} shows the free-fall time, $t_{ff} \equiv
\left ( \frac{3\pi}{32 \, G \, \rho_{\rm av}} \right ) ^{1/2}$, of the
gas in G8.68$-$0.37 as a function of radius, where $\rho_{\rm av}$ is
the average density within each volume calculated from the enclosed
mass within that radius. Taking 10$^4$\,yr (the free-fall time at
0.05\,pc) as a lower limit of the protocluster age and following
\citet{walsh2004}, cores moving 0.01\,pc would reach a velocity of
2.6\,\kms. Future molecular line observations with sufficient angular
resolution to identify MM1 to MM3 should easily be able to discern
such large relative velocities. The inferred proper motions of
$\sim$60\,$\mu \arcsec$/yr are too small to detect.

\subsection{Considering non-thermal fragmentation}
\label{sub:non_thermal_frag}

\begin{figure*}
\begin{center}
\begin{tabular}{c}
\includegraphics[width=12cm, angle=-90]{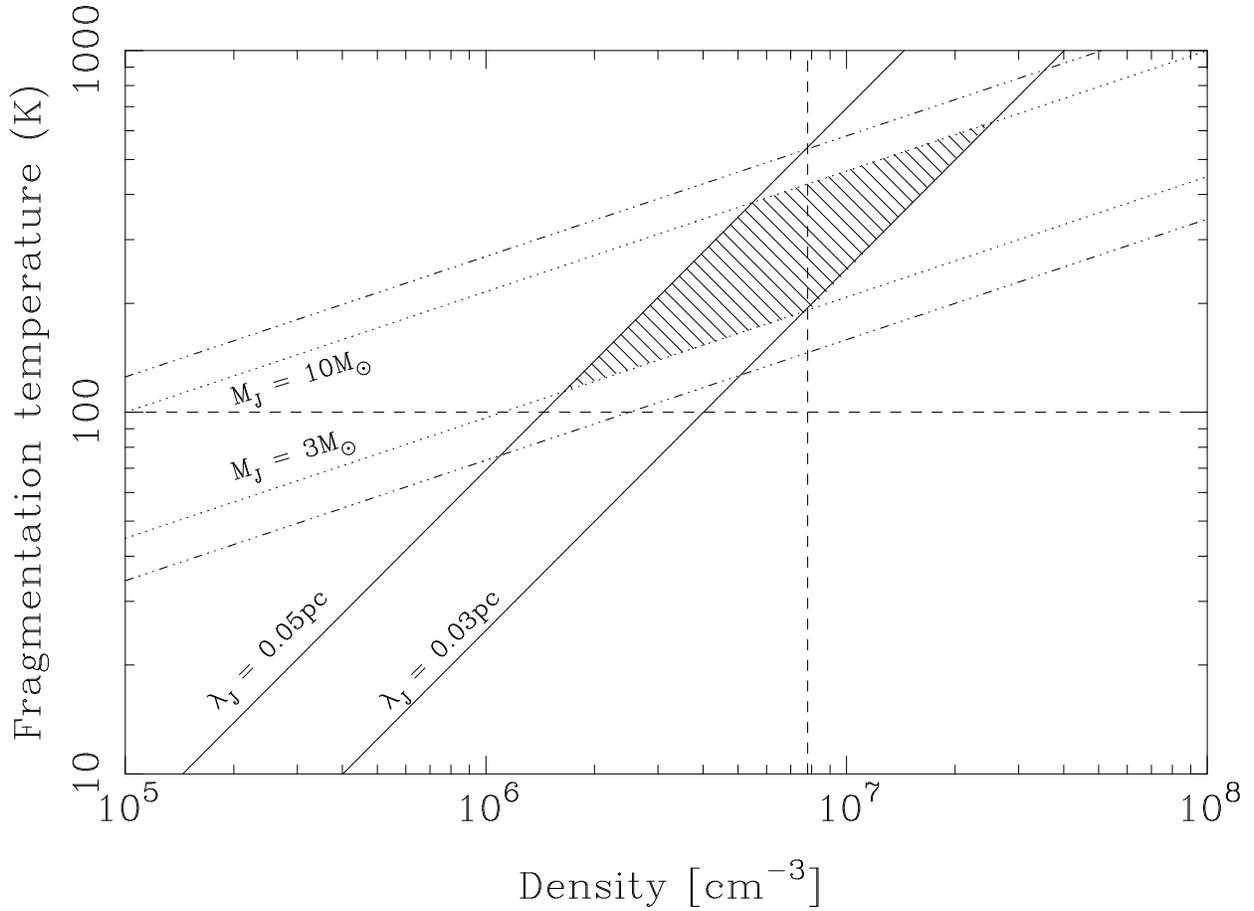} \\
\\
\end{tabular}
\end{center}
\vspace{3mm}
\caption{Encapsulating the observational uncertainties in the measured
  separations and mass of MM1 to MM3 to illustrate how these affect
  the inferred fragmentation temperature. The solid lines show the
  expected T$_{\rm frag}$ as a function of density for Jeans lengths
  of 0.05 and 0.03\,pc, corresponding to upper and lower limit
  estimates of the MM1 to MM3 separations. The dotted lines show the
  expected T$_{\rm frag}$ as a function of density for Jeans masses of
  3 and 10\,M$_\odot$, corresponding to reasonable estimates of the
  MM1 to MM3 masses. The dot-dashed lines illustrate the most extreme
  mass range given the observational uncertainties.  The vertical
  dashed line shows the average density within a volume of
  $r<0.05$\,pc (the radius encompassing MM1 to MM3) of
  7.8$\times$10$^6 \cm3$ as a reasonable estimate for the
  fragmentation density. The shaded region illustrating the
  observational limits shows the fragmentation temperature is at least
  100\,K (the horizontal dashed line). If the density were lower in
  the past (for example if the region were undergoing global infall)
  the fragmentation temperature would also be lower.}
\label{fig:g8.68_tfrag}
\end{figure*}

In the preceding sections we have used Jeans analysis to determine the
mass and separation of core fragments, given the temperature and
density of the surrounding environment. While the assumption of purely
thermal support is undoubtedly an oversimplification -- both
turbulence and magnetic fields are also likely important as support
mechanisms at the earliest stages of massive star formation
\citep[e.g.][]{zhang2009,girart2009,tang2009a,tang2009b,tang2010} --
the relative importance of the different support mechanisms is not
well understood at these spatial scales and for objects at such an
early evolutionary stage. Without a direct measurement of the magnetic
field towards G8.68$-$0.37 we make the assumption of equipartition
between thermal, turbulent and magnetic-field support and assess how
this affects the fragmentation temperature determined in
$\S$~\ref{subsub:static_tfrag}. Returning to
Fig~\ref{fig:g8.68_tfrag}, we used a lower-limit fragmentation
temperature of 100\,K in the analysis of $\S$~\ref{sub:static}.  If
temperature, magnetic fields and turbulence are in equipartition the
total energy is equivalent to a temperature of
300\,K. Fig~\ref{fig:g8.68_tfrag} shows that 300\,K is in fact a more
realistic fragmentation temperature than the 100\,K previously
assumed. We therefore conclude that the results derived assuming
purely thermal fragmentation are robust when including support from
turbulence and magnetic fields.

Finally, we consider the assumption in the Jeans analysis that the gas
is isothermal. \citet{krumholz2007} find in their numerical models
that at high column densities, where the gas becomes optically-thick,
the equation of state deviates from isothermal. The result is that
fragmentation is suppressed at a lower temperature than would be
expected from Jeans analysis. While this is difficult to test with
observations of a single region, in future work we will look to
compare the mass distribution of cores above and below the critical
column density of ${\rm 1\,g\,cm^{-2}}$, at which this affect is
predicted to become important \citep{krumholz_mckee2008}.

\subsection{Will the G8.68$-$0.37 protocluster form O stars?}
\label{sub:mm1_to_mm3_mass}

In order to form an O star through direct collapse, a fragment would
need to have a mass of at least the stellar mass times by the star
formation efficiency -- ie of order 100\,M$_\odot$ of gas, given
typical star formation efficiencies \citep[e.g.][]{ladalada2003}.  No
matter what the fragmentation history of the region, none of the
observed MM1 to MM3 cores has this much mass. Any feedback (if it
exists/existed) was therefore not sufficient to create a fragment that
can collapse to form an O star. In order to form an O star via
accretion, these existing cores need to obtain additional mass from
elsewhere. We now investigate whether G8.68$-$0.37 is likely to form O
stars and, if so, how this might occur.

The total mass within $r<0.05$\,pc (encompassing MM1 to MM3) predicted
from the best-fit model density profile is $\sim$100\,M$_\odot$. MM1
to MM3 account for roughly a third of the mass, leaving a large
reservoir of material inside this relatively small volume. If the
measured infall rate (see $\S$~\ref{subsub:global_infall}) can be
sustained for a few 10$^5$\,yr and large-scale infall can feed mass to
smaller spatial scales \citep[as observed towards G20.08$-$0.14N
  by][]{galvan_madrid_2009}, it seems plausible that MM1 to MM3 may
end up as O stars via continued accretion. However, without
information of the gas velocities at $r < 0.05$\,pc, it is not clear
how MM1 to MM3 are coupled to the gas at this size scale and whether
or not they will continue to gain mass. Several theoretical scenarios
are postulated: continued accretion through a common rotating envelope
or `disk' \citep[e.g.][]{keto2003, keto2007, krumholz2009}; accretion
from initially unbound cluster-scale gas \citep[see
  e.g.][]{smith2009}, or fragmentation-induced starvation via
gravitational instabilities in the accretion flow \citep{peters2010}.

Considering the first scenario -- that MM1 to MM3 may be fed by
accretion through a common rotating envelope or disk -- we do not
detect molecular-line emission in the highest angular resolution
observations so can not search for kinematic signatures of rotation at
this spatial scale. However, it is interesting to note that MM1 to MM3
are oriented roughly in a plane perpendicular to the outflow. This is
similar to the results of \citet{fallscheer2009} towards 18223$-$3 --
another young high mass protocluster, with similar global mass, infall
and outflow rates. They find the mm continuum emission peaks are
aligned perpendicular to the outflow and encompassed by a flattened,
rotating entity of inward spiralling molecular gas. Although emission
from a rotating envelope of 28,000\,AU would have been resolved in the
observations of G8.68$-$0.37, it is not seen. But such a large
rotating envelope is not necessarily expected. A recent large survey
to find high mass accretion disk candidates \citep{bwl2009} found
diameters of the rotating entities to be $\sim$10,000\,AU, smaller
than the unresolved bright molecular line emission in the SMA
observations in this work.

The second scenario is similar to that described in
$\S$~\ref{subsub:sf_on_the_move} where the cores initially form with
smaller mass and at larger radii. Over time they move closer under the
influence of the global gravitational potential. In this picture,
cores grow by accreting the diffuse surrounding unbound cluster-scale
gas. With this in mind, it is interesting that the most massive
observed mm continuum core, MM1, is found at the center of the
gravitational potential (as predicted) where the infall would be
greatest. However, other than searching for the relative velocity of
the cores (see $\S$~\ref{subsub:sf_on_the_move}) it is difficult to
test these predictions with observations of a single region. We are
currently in the process of extending this analysis to larger samples
of massive star formation regions \citep[e.g. from the $\nhthree$
  cores detected in the HOPS Galactic Plane survey][]{walsh2008}.

In summary, we conclude that G8.68$-$0.37 may still proceed to form O
stars, with large scale infall and continued accretion at smaller
scales feeding the observed mm cores. Observations to resolve the gas
kinematics at $r < 0.05$\,pc are required to determine i) how the mm
cores and remaining gas are coupled at this size scale and ii) the
mechanism through which accretion is taking place.

\section{Conclusions}
\label{sec:conc}

Combining multiple-configuration SMA continuum observations with
modelling of the $\nhthree$ and $\methylcyanide$ emission, we have
determined the present-day i) temperature/density as a function of
radius and, ii) mass/separation of cores, in the massive protocluster
G8.68$-$0.37. We then used this investigate whether feedback from low
mass stars can raise the protocluster gas temperature sufficiently to
delay thermal fragmentation, allowing massive stars to form through
direct collapse of high-mass fragments.

From radii of 0.58\,pc (1.2$\times$10$^5$\,AU) down to 0.05\,pc
(10$^4$\,AU) we find the data are well fit with the region having a
power law temperature and density of the form T$ \propto r^{-0.35}$
and $\rho \propto r^{-2.08}$. At $r<0.05$\,pc the assumption of spherical
symmetry breaks down and the temperature/density range are calculated
to be 100$-$200\,K and 10$^{6}-10^{7}\cm3$, respectively. Within this
radius of 0.05\,pc, the SMA resolves the 1.2\,mm continuum emission
into 3 cores, MM1 to MM3, with separations of 6200\,AU and 9700\,AU
and masses 10$\pm$3\,M$_\odot$, 7$\pm$2\,M$_\odot$ and
4$\pm$2\,M$_\odot$, respectively.

From the region's observed properties, we infer the conditions at the
time the cores formed by fragmentation. Assuming the measured
separations of MM1 to MM3 and the average density encompassing these
cores ($r<0.05$\,pc) are representative of the physical conditions at
fragmentation, this implies a thermal fragmentation temperature of at
least 100\,K.  We rule out a population of low mass stars being able
to provide this heating -- an unfeasibly large number are required and
the measured bolometric luminosity is orders of magnitude too small
even allowing for heating due to trapped accretion luminosity. The
heating could instead be provided by the equivalent of a 5 M$_\odot$
ZAMS star but this is an unsatisfactory source of thermal feedback --
it invokes a circular argument requiring the star to have raised the
temperature to suppress fragmentation, before it
formed. Alternatively, the required fragmentation temperature could be
lower if the region were initially at a lower density, for example if
the cores formed farther apart or the region was undergoing global
infall.

Whatever the fragmentation history, none of the observed cores has
sufficient raw material to form an O star through direct collapse.
Even if feedback may have suppressed fragmentation, it was not
sufficient to halt it to this extent. If G8.68$-$0.37 is destined to
form O stars, the observed cores must obtain additional mass from
outside their observationally derived boundaries. The observations
suggest that the cores in this protocluster are being fed via global
infall from the very massive reservoir ($\sim$1500\,M$_\odot$) of gas
within which the protocluster is embedded.

\section{Acknowledgements}

We would like to thank the anonymous referee for their thoughtful and
constructive comments. SNL would like to thank Stella Offner, Phil
Myers, David Wilner, Tyler Bourke, Rowan Smith and Paul Ho for
thoughtful comments, Roberto Galvan-Madrid for comments on a draft of
the paper and Mark Krumholz for a stimulating discussion on the
analysis. SNL gratefully acknowledges support of this research through
funding as a Submillimeter Array Fellow. The Submillimeter Array is a
joint project between the Smithsonian Astrophysical Observatory and
the Academia Sinica Institute of Astronomy and Astrophysics and is
funded by the Smithsonian Institution and the Academia Sinica. This
research has made use of NASA's Astrophysics Data System.

\bibliography{g8_68_sma_bib}


\begin{table*}
    \caption{Overview of the observing setups. The first two columns
      show the array configuration and observing frequencies. The
      sub-compact, compact and extended array configurations have
      baselines ranging from 31 - 89\,m, 61 - 192\,m and 82 - 247\,m,
      respectively. The third and final columns give the continuum
      sensitivity and synthesised beam size.  }
  \begin{center}
    \begin{tabular}{|c|c|c|c|}\hline\hline      
 Array         & Freq.      & Cont. RMS        &  Beam \\ 
 Config.       & (GHz)     & (mJy/beam)       & ($\arcsec$)   \\ \hline
 Sub-compact  & 279       &  7.3             &  7.4$\times$6.2 \\
 Compact       & 230       &  1.4             &  3.1$\times$2.0 \\ 
 Extended      & 217       &  3.4             &  1.1$\times$1.0 \\\hline  

    \end{tabular}

    \label{tab:obs_setup}
  \end{center}
  \vspace{-2mm}
\end{table*}

\begin{table*}
    \caption{Properties of the mm continuum sources detected with the
      SMA in the extended array configuration.  }
  \begin{center}
    \begin{tabular}{|c|c|c|c|c|}\hline\hline      
Source & Flux (mJy) & Flux Peak (mJy) & R.A.          &  Dec.        \\ \hline
MM1    & 45     & 37          & 18:06:23.479  & -21:37:10.5 \\
MM2    & 30     & 24          & 18:06:23.522  & -21:37:11.7 \\
MM3    & 17     & 14          & 18:06:23.494  & -21:37:09.1  \\ \hline

    \end{tabular}

    \label{tab:mm_src_props}
  \end{center}
  \vspace{-2mm}
\end{table*}

\begin{table*}
    \caption{$\nhthree$ radiative transfer modelling parameters. A
      spherically-symmetric model of radius 0.58\,pc was chosen based
      on the extent/morphology of the $\nhone$ emission. The
      temperature, T$_K$, and density, $\rho$, were parameterized with
      radius, $r$, as power laws:
      $\rho(r)=\rho_{1/2}\;~r^{-\kappa_\rho}$ and T$_K(r)=$T$_{1/2}\;
      r ^{-\kappa_T}$, where $\rho_{1/2}$ and T$_{1/2}$ are the
      density and temperature at the half radius point, 0.29\,pc. An
      additional constant velocity component, $\Delta V_{\rm NT}$, was
      included to model non-thermal support (e.g. from turbulence),
      resulting in a 5 parameter model for the cloud. The second and
      third columns show the range of parameter space the models
      searched through. The final column gives the best fit model to
      the L07A data after $\chi^2$ minimisation of 10000 models using
      simulated annealing to search through the 5D parameter
      space. Full details of the modelling are given in
      $\S$~\ref{subsub:nh3_modelling}. }
  \begin{center}
    \begin{tabular}{|c|c|c|c|}\hline\hline      
 Parameter                   &  \multicolumn{2}{|c|}{Search range}      &  Best-fit \\ 
                             &  Min  & Max        &  Value        \\ \hline
 log ($\rho_{1/2}$ [$\cm3$])  & 3.0   & 8.0     & 4.98 \\
 $\kappa_\rho$                & 0.5   &   4.0   & 2.08 \\
 T$_{1/2}$ [K]                & 10.0  & 400.0   & 41.5 \\
 $\kappa_T$                  &  0.1  & 2.5     & 0.35 \\
 $\Delta V_{\rm NT}$ [\kms]   & 0.5   & 3.0     & 1.79 \\ \hline
    \end{tabular}

    \label{tab:model_pars}
  \end{center}
  \vspace{-2mm}
\end{table*}


\newpage

\begin{figure*}
\begin{center}
  \begin{tabular}{ccc}
    \includegraphics[width=3.5cm, angle=-90, trim=0 0 -5 0]{pos_0_trans_0_fixed_yaxis.ps}&
    \includegraphics[width=3.5cm, angle=-90, trim=0 0 -5 0]{pos_1_trans_0_fixed_yaxis.ps}&
    \includegraphics[width=3.5cm, angle=-90, trim=0 0 -5 0]{pos_2_trans_0_fixed_yaxis.ps}\\

    \includegraphics[width=3.5cm, angle=-90, trim=0 0 -5 0]{pos_0_trans_3_fixed_yaxis.ps}&
    \includegraphics[width=3.5cm, angle=-90, trim=0 0 -5 0]{pos_1_trans_3_fixed_yaxis.ps}&
    \includegraphics[width=3.5cm, angle=-90, trim=0 0 -5 0]{pos_2_trans_3_fixed_yaxis.ps}\\
 
    \includegraphics[width=3.5cm, angle=-90, trim=0 0 -5 0]{pos_0_trans_4_fixed_yaxis.ps}&
    \includegraphics[width=3.5cm, angle=-90, trim=0 0 -5 0]{pos_1_trans_4_fixed_yaxis.ps}&
    \includegraphics[width=3.5cm, angle=-90, trim=0 0 -5 0]{pos_2_trans_4_fixed_yaxis.ps}\\

 \end{tabular}
\end{center}
\caption{Results of the $\nhthree$ radiative transfer modelling
  outlined in $\S$~\ref{subsub:nh3_modelling}. \citet{L07A} spectra
  are shown in black, overlayed with the best-fit model spectra in
  red. Each column shows spectra taken from a single position on the
  sky.  The top, middle and bottom rows are the $\nhone$, (4,4) and
  (5,5) transitions, respectively. The positions were selected to be
  spaced by roughly one synthesised beam in the direction
  perpendicular to the molecular outflow (see $\S$~\ref{sub:outflow}
  and $\S$~\ref{subsub:nh3_modelling}) to minimize potential
  contamination from gas affected by the outflow. }
\label{fig:g8.68_rt_fit}
\end{figure*}

%

\end{document}